\documentstyle[aps]{revtex}
\draft

\begin{document}
\author{Chueng-Ryong Ji and Yuriy Mishchenko}
\title{The General Theory of Quantum Field Mixing}
\address{Department of Physics, North Carolina State University, Raleigh,
North Carolina 27695-8202}
\date{2/5/2002 }
\maketitle

\begin{abstract}
We present a general theory of mixing for an arbitrary number of fields with
integer or half-integer spin. The time dynamics of the interacting 
fields is solved and the Fock space for interacting fields is explicitly 
constructed. The unitary inequivalence of the Fock space of base (unmixed)
eigenstates and the physical mixed eigenstates
is shown by a straightforward algebraic method for any number of 
flavors in boson or fermion statistics. The oscillation formulas based on 
the nonperturbative vacuum are 
derived in a unified general formulation and then applied to both two and 
three flavor 
cases. Especially, the mixing of spin-1 (vector) mesons and the CKM 
mixing phenomena in the Standard Model are discussed emphasizing the
nonperturbative vacuum effect in quantum field theory.
\end{abstract}

\tightenlines

\baselineskip=20pt \setcounter{section}{0} \setcounter{equation}{0} 
\setcounter{figure}{0} 
\renewcommand{\theequation}{\mbox{1.\arabic{equation}}} 
\renewcommand{\thefigure}{\mbox{1.\arabic{figure}}}

\section{Introduction}

The mixing of quantum fields plays an important role in the phenomenology
of high-energy physics \cite{1,15,18}. 
Mixings of both $K^0\bar K^0$ and $B^0\bar B^0$
bosons provide the evidence of $CP$ violation in the weak interaction \cite{14}
and $\eta\eta \prime $ boson mixing in the $SU(3)$ flavor group gives a 
unique opportunity to investigate the nontrivial QCD vacuum and fill 
the gap between QCD and the constituent quark model. In the fermion 
sector, neutrino mixing and oscillations are the likely resolution of the 
famous solar neutrino puzzle \cite{16,17,26}. In addition, the standard model 
incorporates the mixing of fermion fields through the 
Kobayashi-Maskawa (CKM) mixing of three quark flavors, a generaliztion 
of the original Cabibbo mixing matrix between the $d$ and $s$ quarks
\cite{11,22,24,29}.
Therefore, careful theoretical analyses of the mixing
problem in quantum field theory is an important step toward understanding
the many-body aspects of high-energy phenomena and their relationship to
other areas of physics involving phase transitions. 

Moreover, the theory of mixing fields touches important, yet not fully
answered, fundamental question about the quantization of the interacting
fields. The mixing transformation introduces very non-trivial
relationships between the interacting and non-interacting (free) fields,
which lead to a unitary inequivalence between the two Fock spaces \cite{3,4} 
of the interacting fields and the free fields. 
This is different
from the perturbation theories where the vacuum state of interacting
fields is equal to the vacuum of free fields up to a less essential phase 
factor $e^{iS_0}$ \cite{10,12,21}. The mixing of quantum fields is one of the cases 
that can be solved nonperturbatively in the quantum
field theory. Thus, it also allows to investigate the accuracy of 
perturbation theory. For instance, the dynamics of a mixed-field 
Hamiltonian can be used for a partial summation of regular perturbation 
series as well as an improvement of the
accuracy in perturbation theory. 

Recently, importance of the mixing transformations has prompted
a fundamental examination of them from a quantum field theoretic perspective.
The investigation of two-field unitary mixing in the fermion case 
demonstrated a rich structure of the interacting-field vacuum as SU(2) 
coherent state and altered the oscillation formula including the 
antiparticle degrees of freedom. 
Momentum dependence of mixing, existence
of correlated antiparticle beam and additional high-frequency
oscillation terms have been found and at the same time the vacuum 
condensates have been analyzed for fermions \cite{4,5,8,9,19,20}. 
Subsequent analyses for the boson case revealed similar
features but much more complicated vacuum structure for 
interacting fields \cite{3,7,23}. Especially, the pole structure in the inner 
product between the mass vacuum and the flavor vacuum was found and 
related to the convergence limit of perturbation series \cite{23}. 
Attempts to look at the mixings of three-fermion
case have also been carried out \cite{4,8,20}.

In this paper, we extend the previous analyses of mixing phenomena and 
work out a unified theoretical framework for an arbitrary number of 
flavors with any integer (bosons) or half-integer (fermions) spin 
statistics. We build the representation of mixing transformation in
the Fock space of quantum fields and demonstrate how this can be used 
to obtain exact oscillation effects. We then use the developed 
framework to carry out calculations of two-field and three-field unitary 
mixings for the typical spin (i.e. 0,1/2 and 1) cases. 
We also comment on the use of mixed-field solution to improve the 
perturbation series of mixing effects.

The paper is organized as follows. In Section II, we define the ladder 
operators for flavor fields and explicitly show the unitary inequivalence 
between the flavor Fock space and the Fock space of mass-eigenstates. In 
Section III, we find the time dynamics of the flavor ladder operators and 
derive general expressions for the particle condensations and the number 
operators as functions of time. We also present some remarks on Green 
function method in the mixing problem. We then specifically consider, in 
Section IV, the mixing of two spin-1 fields (vector mesons) along with
the mixing of spin-1/2 fields and show the consistency with previously known 
results. 
Summary and conclusion follow in sections V.
In Appendix A, the mixing parameters are shown explicitly for the spin 0, 
1/2 and 1. In Appendix B, we present a derivation of the 
flavor vacuum state by solving an infinite system of
coupled equations which appears as a condition of the vacuum annihilation. 
In Appendix C, we summarize our results of the three-field mixing for the 
spin 0,1/2 and 1 using the SU(3) Wolfenstein parametrization. 

\setcounter{equation}{0} \setcounter{figure}{0} 
\renewcommand{\theequation}{\mbox{2.\arabic{equation}}} 
\renewcommand{\thefigure}{\mbox{2.\arabic{figure}}}

\section{The Theory of Quantum Field Mixings}

In this section, we consider the mixing problem for 
$N$ fields of fermions or bosons.
To discuss the dynamics of the flavor (mixed) 
fields, we define a flavor field $\phi_{\mu}$ as 
a mixture of the free fields $\varphi_{j}$ ($j=1,2,\ldots,N$); i.e. 
\begin{equation}
\label{mixing}\phi _\mu =\sum_jU_{\mu j}\varphi _j,
\end{equation}
where $U_{\mu j}$ is a unitary mixing matrix element. 
We use the latin indices $i,j,k,\ldots $ to label the fields
of mass-eigenstates and the greek indices 
$\mu ,\nu ,\xi ,\ldots $ to label the flavor fields. 
We also denote $\bar \phi$ and $\bar \varphi$ as
the entire columns
$\bar \phi =(\phi_1,\phi_2\ldots,\phi_N)^{\top}$ and
$\bar \varphi =(\varphi _1,\varphi _2\ldots,\varphi _N)^{\top }$,
respectively.
The evolution of the fields $\phi _{\mu}$ 
is generated by the Hamiltonian of the form 
\footnote{When there is an additional interaction Hamiltonian for 
$\bar \phi$ given by $H_I=\bar \phi^{\dagger}W {\bar\phi}$,
the Hamiltonian of the system is, of course, extended to
$\tilde{H}(\bar \phi)=H(\bar\phi)+H_I=H_0(\bar\phi)+
\bar\phi^{\dagger}{\cal M}\bar\phi+\bar\phi^{\dagger}W{\bar\phi}$.
Then $H_{free}(\bar\varphi)$ is also extended to 
$H_{free}(\bar\varphi)+\bar\varphi^{\dagger}U^{\dagger}WU\bar\varphi$.}
\begin{equation}\label{hamiltonians}
H(\bar \phi )=H_{free}(\bar\varphi)=
H_{free}(U^{\dagger}\bar\phi)=H_{0}(\bar\phi)+\bar\phi^{\dagger }
{\cal M}\bar\phi,
\end{equation}
where $H_{free}(\bar\varphi)$ is the free field Hamiltonian
for $\varphi_{i}$ with the corresponding mass eigenvalues $m_i$, 
$H_{0}(\bar\phi)$ is the free flavor field Hamiltonian and ${\cal M}$ is 
a mixing matrix. 

The existence of the explicit relationship between free ($\varphi $) and flavor 
($\phi $) fields, given by Eq.(\ref{mixing}), 
allows us to work out the quantum-field theoretical solution
to the problem given by 
\begin{equation}\label{mixingproblem}
\frac d{dt}\phi _\mu =i[H(\bar\phi),\phi _\mu ].
\end{equation}
In fact, the solution of Eq.(\ref{mixingproblem}) is 
contained in Eq.(\ref{mixing}) with 
the free field ($\varphi _i$) given by
\begin{equation}\label{freefield}
\varphi _i=\sum_\sigma 
\displaystyle \int 
\frac{d\vec k}{\sqrt{2\epsilon _{i\vec k}}}\left( u_{\vec k\sigma }^ia_{i\vec 
k\sigma}\left( t\right) e^{i\vec k\vec x}+v_{\vec k\sigma }^ib_{i\vec k
\sigma}^{\dagger }\left( t\right) e^{-i\vec k\vec x}\right) ,
\end{equation}
where $a_{i\vec k\sigma}(t)=e^{-i\epsilon _{i\vec k}t}a_{i\vec 
k\sigma}$ and
$b_{i\vec k\sigma}(t)=e^{-i\epsilon _{i\vec k}t}b_{i\vec k\sigma}$ 
with the standard equal time commutation/anticommutation relationships
for bosons/fermions, i.e. 
$$
[a_\alpha \left( t\right) ,a_{\alpha ^{\prime }}^{\dagger }\left( t\right)
]_{\pm }=[b_\alpha \left( t\right) ,b_{\alpha ^{\prime }}^{\dagger }\left(
t\right)]_{\pm}=\delta _{\alpha ,\alpha ^{\prime }}.
$$
In Eq.(\ref{freefield}), $u^i_{\vec k\sigma }$ and $v^i_{\vec k\sigma }$ are 
the free particle and antiparticle
amplitudes, respectively, and $\sigma $ is the helicity quantum number
given by 
\begin{equation}
\label{chirality}
\left( \vec n\cdot \vec s\right) u_{\vec k\sigma }^i=
\sigma u_{\vec k\sigma }^i ,
\left( \vec n\cdot \vec s\right) v_{\vec k\sigma }^i=
\sigma v_{\vec k\sigma }^i,
\end{equation}
where $\vec s$ is the spin operator and $\vec n=\vec k/|\vec{k}|$.
We also define the following parameters that are useful in 
extracting the ladder operators from the field operators
\begin{equation}
\label{Hhparams}
\begin{array}{c}
H_{
\vec k\sigma }^{\mu j}\delta _{\sigma ,\sigma ^{\prime }}=u_{\vec k\sigma
}^{\mu \dagger }u_{\vec k\sigma ^{\prime }}^j=v_{-\vec k-\sigma }^{\mu
\dagger }v_{-\vec k-\sigma ^{\prime }}^j, \\ h_{\vec k\sigma }^{\mu j}\delta
_{\sigma ,\sigma ^{\prime }}=u_{\vec k\sigma }^{\mu \dagger }v_{-\vec k
-\sigma ^{\prime }}^j.
\end{array}
\end{equation}
For the analysis of arbitrary flavor mass parametrizations, we use the 
general notation given by Eq.(\ref{Hhparams}) including both flavor and 
mass degrees of freedom. Although both indices $\mu$ and $j$ are numbers
running from 1 to $N$, the mass for the first index should be 
used as the flavor mass while the second index is for the mass eigenvalue 
$m_j$. One should note that $H$ and $h$ are both
symmetric for bosons while $H$ is symmetric and $h$ is antisymmetric
for fermions. The explicit representations of $H$ and $h$ are presented in the Appendix
A for the spin 0,1/2 and 1 cases.

Now, if $\Lambda (U,t)$ is the representation of the mixing transformation defined in
the equal time quantization, then 
\begin{equation}
\label{mixrepresentation}
\bar \phi (t)=U\bar \varphi (t)=
    \Lambda (U,t)^{\dagger }\bar \varphi(t)\Lambda (U,t).
\end{equation}
In the associate Fock-space, this corresponds to
\begin{equation}\label{mixrepresentation1}
|\alpha ,t>_f =\Lambda (U,t)^{\dagger }|\alpha ,t>_m
\end{equation}
where subscript $f$ ($m$) is used to denote the flavor
(mass) Fock-space. For the given time $t$, Eq.(\ref{hamiltonians}) can
then be written as 
\begin{equation}
H(\bar\phi(t))=\Lambda (U,t)^{\dagger }H_{free}(\bar\varphi(t))\Lambda (U,t).
\end{equation}
As noticed from the two-field mixing analysis \cite{3,4,7,23}, 
$H(\bar \phi(t))$ and $H(\bar \varphi (t))$ cannot be in general 
related by the same operator at all times so that $\Lambda (U,t)$ 
is essentially time dependent. The
vacuum state of the flavor-fields, defined as the state with the minimum
energy, is $\Lambda (U,t)^{\dagger }|0>_m$ and changes with time satisfying 
\begin{equation}
_f<\alpha |H(\bar\phi(t))|\alpha >_f=
_m<\alpha |H(\bar\varphi(t))|\alpha >_m\geq _m<0|H(\bar\varphi(t))|0>_m=
_f <0|H(\bar\phi(t))|0>_f.
\end{equation}
We now define the ladder operators for the flavor fields as 
$\tilde a_{\mu=i,\vec k\sigma}(t)=
\Lambda (U,t)^{\dagger }a_{i\vec k\sigma }(t)\Lambda (U,t)$. 
Using linearity of the mixing transformation, we then can solve
the explicit structure of $\tilde a_{\mu \vec k\sigma }(t)$ without finding 
$\Lambda (U,t)$ itself.

Such approach in fact has been known for some time for the fermion case \cite{20},
where it was noticed that fermion ladder operators for spin 1/2 can be
extracted from quantum fields by means of 
\begin{equation}
\label{ladfermion}
\begin{array}{l}
a_{i{\vec k}\sigma }(t)=\frac{\sqrt{2\epsilon _{i{\vec k}}}}{H_{\vec k\sigma 
}^{ii}}u_{
\vec k\sigma }^{i\dagger }\varphi _{i\vec k}(t), \\ b_{i-\vec k-\sigma
}(t)=\left[ \frac{\sqrt{2\epsilon _{i\vec k}}}{H_{\vec k\sigma 
}^{ii}}v_{-\vec k
-\sigma }^{i\dagger }\varphi _{i\vec k}(t)\right] ^{\dagger }.
\end{array}
\end{equation}
Since the Fourier component $\varphi _{i\vec k}(t) = \sum\limits_\sigma 
\frac{1}{\sqrt{\epsilon_{i\vec k}}} \left ( u^i_{{\vec k} \sigma} a_{i{\vec 
k}\sigma}(t) + v^i_{-{\vec k} \sigma} b^\dagger_{i -{\vec k} 
\sigma}(t) \right )$ is obviously a linear
combination of $\varphi _i\left( \vec x,t\right) $, one can express ladder
operators as linear combinations of the initial fields. Using the linearity
of Eq.(\ref{mixrepresentation}), we get 
\begin{equation}
\label{eq01x01}
\begin{array}{c}
\tilde a_{\mu\vec k\sigma }(t)=
\frac{\sqrt{2\epsilon _{\mu \vec k}}}{H_{\vec k\sigma}^{\mu\mu}}u_{\vec 
k\sigma }^{\mu\dagger }
(\Lambda (U,t)^{\dagger }\bar \varphi _{\vec k}(t)\Lambda (U,t))_\mu= 
\\ =\sum\limits_j\frac{\sqrt{2\epsilon _{\mu \vec k}}}{H_{\vec k\sigma 
}^{\mu\mu}} u_{\vec k\sigma}^{\mu\dagger }U_{\mu j}\varphi _{j\vec k}(t), 
\\ \tilde b_{\mu-\vec k-\sigma }(t)=\sum\limits_j\frac{\sqrt{2\epsilon 
_{\mu \vec k}}}{H_{\vec k\sigma }^{\mu\mu}}U_{\mu j}^{*}\varphi _{j\vec k}^
{\dagger }(t)v_{-\vec k-\sigma }^\mu.
\end{array}
\end{equation}

For the bosons, however, the ladder
operators are not separated as in the fermion case, e.g. 
\begin{equation}\label{eq013}
u_{\vec k\sigma }^{i\dagger }\varphi _{i\vec k}(t)=
\frac 1{\sqrt{2\epsilon _{i\vec k}}}(a_{i\vec k\sigma }(t)+
h_{\vec k\sigma }^{ii}b_{i-\vec k-\sigma }^{\dagger }(t))
\end{equation}
and in general $h_{\vec k\sigma }^{ii}\neq 0$. Eq.(\ref{eq013}) implies
that particles and antiparticles in boson case can not be distinguished
unless time dynamics is considered. To deal with this problem we define
ladder operators for bosons by
\begin{equation}
\label{ladboson}
\begin{array}{c}
a_{i
\vec k\sigma }=u_{\vec k\sigma }^{i\dagger }\left( \sqrt{\frac{\epsilon 
_{i \vec k}}2}
\varphi _{i\vec k}(t)+\frac 1{\sqrt{2\epsilon _{i \vec k}}}\dot \varphi 
_{i\vec k}(t)\right) , 
\\ b_{i-\vec k-\sigma }^{\dagger }=v_{-\vec k-\sigma
}^{i\dagger }\left( \sqrt{\frac{\epsilon _{i \vec k}}2}\varphi _{i\vec 
k}(t)-\frac 1{\sqrt{2\epsilon _{i \vec k}}}\dot \varphi _{i\vec 
k}(t)\right) . 
\end{array}
\end{equation}
With Eqs.($\ref{ladfermion}$) and ($\ref{ladboson}$), we then derive for
fermions\footnote{Here, we abbreviate the notations $a_{j{\vec k}\sigma}$
and $b^\dagger_{j-{\vec k}-\sigma}$ as $a_j$ and $b^\dagger_{-j}$, 
respectively. Similar abbreviation is used for $\tilde a_\mu$ and $\tilde 
b_{-\mu}$.}: 
\begin{equation}
\label{la01}
\begin{array}{c}
\tilde a_\mu =\frac{\sqrt{2\epsilon _\mu }}{H^{\mu \mu }}\sum\limits_{j,
\sigma ^{\prime }}(u_{\vec k\sigma }^{\mu \dagger }u_{\vec k\sigma ^{\prime
}}^ja_j+u_{\vec k\sigma }^{\mu \dagger }v_{-\vec k-\sigma ^{\prime
}}^jb_{-j}^{\dagger })\frac{U_{\mu j}}{\sqrt{2\epsilon _j}}= \\ 
=\sum\limits_j\left( \sqrt{\frac{\epsilon _\mu }{\epsilon _j}}\frac{H^{\mu j}
}{H^{\mu \mu }}U_{\mu j}a_j+\sqrt{\frac{\epsilon _\mu }{\epsilon _j}}\frac{
h^{\mu j}}{H^{\mu \mu }}U_{\mu j}b_{-j}^{\dagger }\right) ;
\end{array}
\end{equation}
\begin{equation}
\label{la02}
\begin{array}{c}
\tilde b_{-\mu }=\frac{\sqrt{2\epsilon _\mu }}{H^{\mu \mu }}
\sum\limits_{j,\sigma ^{\prime }}\left( \left( v_{-\vec k-\sigma }^{\mu
\dagger }u_{\vec k\sigma ^{\prime }}^j\right) ^{*}a_{j}^{\dagger
}+\left( v_{-\vec k-\sigma }^{\mu \dagger }v_{-\vec k-\sigma ^{\prime
}}^j\right) ^{*}b_{-j}\right) \frac{U_{\mu j}^{*}}{\sqrt{2\epsilon
_j}}= \\ =\sum\limits_j\left( \sqrt{\frac{\epsilon _\mu }{\epsilon _j}}\frac{
\left( H^{\mu j}\right) ^{*}}{H^{\mu \mu }}U_{\mu j}^{*}b_{-j}-\sqrt{\frac{
\epsilon _\mu }{\epsilon _j}}\frac{\left( h^{\mu j}\right) ^{*}}{H^{\mu \mu }
}U_{\mu j}^{*}a_j^{\dagger }\right) 
\end{array}
\end{equation}
and for bosons: 
\begin{equation}
\label{la03}
\begin{array}{c}
\tilde a_\mu =\frac{\sqrt{2\epsilon _\mu }}2\sum\limits_{j,\sigma ^{\prime
}}\left( u_{\vec k\sigma }^{\mu \dagger }u_{\vec k\sigma ^{\prime }}^j\frac{
\epsilon _\mu +\epsilon _j}{\epsilon _\mu }a_j+u_{\vec k\sigma }^{\mu
\dagger }v_{-\vec k-\sigma ^{\prime }}^j\frac{\epsilon _\mu -\epsilon _j}{
\epsilon _\mu }b_{-j}^{\dagger }\right) \frac{U_{\mu j}}{\sqrt{2\epsilon _j}}
= \\ =\sum\limits_j\left( \frac{\sqrt{\frac{\epsilon _\mu }{\epsilon _j}}+
\sqrt{\frac{\epsilon _j}{\epsilon _\mu }}}2H^{\mu j}U_{\mu j}a_j+\frac{\sqrt{
\frac{\epsilon _\mu }{\epsilon _j}}-\sqrt{\frac{\epsilon _j}{\epsilon _\mu }}
}2h^{\mu j}U_{\mu j}b_{-j}^{\dagger }\right) ;
\end{array}
\end{equation}
\begin{equation}
\label{la04}
\begin{array}{c}
\tilde b_{-\mu }=\frac{\sqrt{2\epsilon _\mu }}2\sum\limits_{j,\sigma
^{\prime }}\left( v_{-\vec k-\sigma }^{\mu \dagger }u_{\vec k\sigma ^{\prime
}}^j\frac{\epsilon _\mu -\epsilon _j}{\epsilon _\mu }a_j^{\dagger }+v_{-\vec k
-\sigma }^{\mu \dagger }v_{-\vec k-\sigma ^{\prime }}^j\frac{\epsilon _\mu
+\epsilon _j}{\epsilon _\mu }b_{-j}\right) ^{*}\frac{U_{\mu j}^{*}}{\sqrt{
2\epsilon _j}}= \\ =\sum\limits_j\left( \frac{\sqrt{\frac{\epsilon _\mu }{
\epsilon _j}}+\sqrt{\frac{\epsilon _j}{\epsilon _\mu }}}2(H^{\mu
j})^{*}U_{\mu j}^{*}b_{-j}+\frac{\sqrt{\frac{\epsilon _\mu }{\epsilon _j}}-
\sqrt{\frac{\epsilon _j}{\epsilon _\mu }}}2(h^{\mu j})^{*}U_{\mu
j}^{*}a_j^{\dagger }\right) .
\end{array}
\end{equation}
Denoting the spin of the mixed fields as $S$, we 
can unify the expressions for both fermion and boson 
in an identical form as 
\begin{equation}
\label{ladshort}
\begin{array}{l}
\tilde a_\mu =\sum\limits_j\left( \alpha _{\mu j}a_j+\beta _{\mu 
j}b_{-j}^{\dagger }\right) , \\ 
\tilde b_{-\mu }=\sum\limits_j\left( \alpha _{\mu
j}^{*}b_{-j}+(-1)^{2S}\beta _{\mu j}^{*}a_j^{\dagger }\right) ,
\end{array}
\end{equation}
by defining 
\begin{equation}
\label{ladshort02}\alpha _{\mu j}=\gamma _{\mu j}^{+}U_{\mu j},\,\beta _{\mu
j}=\gamma _{\mu j}^{-}U_{\mu j},
\end{equation}
where
\begin{equation}
\label{ladshort01}
\begin{array}{c}
\gamma _{\mu j}^{+}=\left\{ 
\begin{array}{c}
\sqrt{\frac{\epsilon _\mu }{\epsilon _j}}\frac{H^{\mu j}}{H^{\mu \mu }}\text{
fermions,} \\ H^{\mu j}\frac{\sqrt{\frac{\epsilon _\mu }{\epsilon _j}}+\sqrt{
\frac{\epsilon _j}{\epsilon _\mu }}}2\text{ bosons.}
\end{array}
\right.  \\ 
\gamma _{\mu j}^{-}=\left\{ 
\begin{array}{c}
\sqrt{\frac{\epsilon _\mu }{\epsilon _j}}\frac{h^{\mu j}}{H^{\mu \mu }}\text{
fermions,} \\ h^{\mu j}\frac{\sqrt{\frac{\epsilon _\mu }{\epsilon _j}}-\sqrt{
\frac{\epsilon _j}{\epsilon _\mu }}}2\text{ bosons.}
\end{array}
\right. 
\end{array}
\end{equation}
We also note from unitarity that 
\begin{equation}
\left\{ 
\begin{array}{l}
\left| \alpha _{\mu j}\right| ^2+\left| \beta _{\mu j}\right| ^2=\left|
U_{\mu j}\right| ^2
\text{, fermions;} \\ \left| \alpha _{\mu j}\right| ^2-\left| \beta
_{\mu j}\right| ^2=\left| U_{\mu j}\right| ^2\text{, bosons}
\end{array}
\right. 
\end{equation}
so that one can treat $\alpha _{\mu j}$ and $\beta _{\mu j}$ as cosine and sine for
fermions (cosh and sinh for bosons), respectively: 
\begin{equation}
\label{eq03x01}
\begin{array}{c}
\alpha _{\mu j}=U_{\mu j}\left\{ 
\begin{array}{c}
\cos (\theta _{\mu j})
\text{ fermions} \\ \cosh (\theta _{\mu j})\text{ bosons}
\end{array}
\right. , \\ 
\beta _{\mu j}=U_{\mu j}\left\{ 
\begin{array}{c}
\sin (\theta _{\mu j})
\text{ fermions} \\ \sinh (\theta _{\mu j})\text{ bosons}
\end{array}
\right. .
\end{array}
\end{equation}
From the fact that Eqs.(\ref{la01})-(\ref{la04}) serve as the mixing group 
representation, one can conclude that 
\begin{equation}
\label{eq02x01}\theta _{\mu j}-\theta _{\mu j^{\prime }}=\theta _{j^{\prime
}j}
\end{equation}
regardless of $m_\mu $. Using the formulas, presented in the Appendix A, 
this can be explicitly verified for $S=0,1/2,1$ by calculating, 
for example, 
$\frac{
\partial \gamma _{\mu j}^{-}}{\partial m_\mu }$. In every case $\frac
{\partial \gamma _{\mu j}^{-}}{\partial m_\mu }$ can be reduced to $\frac
{\partial \gamma _{\mu j}^{-}}{\partial m_\mu }=\gamma _{\mu j}^{+}
f(m_\mu )$, e.g. for fermions
$$
\frac{\partial \theta _{\mu j'}}{\partial m_\mu }-
\frac{\partial \theta _{\mu j}}{\partial m_\mu }=
\frac{\frac{\partial \sin(\theta _{\mu j'})}{\partial m_\mu }}{\cos (\theta _{\mu j'})}-
\frac{\frac{\partial \sin(\theta _{\mu j})}{\partial m_\mu }}{\cos (\theta _{\mu j})}=
f(m_\mu )-f(m_\mu )=0 
$$
so that $\theta _{\mu j}=\theta _\mu -\theta _j$, where
$\cos(\theta_{\mu})=\frac{1}{2\sqrt{\epsilon_{\mu}}}
(\sqrt{\epsilon_{\mu}+m_{\mu}}+\sqrt{\epsilon_{\mu}-m_{\mu}})$ 
and $\sin(\theta_{\mu})=\frac{1}{2\sqrt{\epsilon_{\mu}}}
(\sqrt{\epsilon_{\mu}+m_{\mu}}-\sqrt{\epsilon_{\mu}-m_{\mu}})$.

The introduced ladder operators are consistent with the representation of the
mixing transformation in the Fock space: 
\begin{equation}
|\alpha _\mu+1,t>_f =\tilde a_\mu^{\dagger }(t)|\alpha _\mu,t>_f =
\Lambda(U,t)^{\dagger }a_i^{\dagger }(t)\Lambda (U,t)
\Lambda (U,t)^{\dagger}|\alpha _i,t>_m=
\Lambda (U,t)^{\dagger }|\alpha _i+1,t>_m,
\end{equation}
and the flavor vacuum state satisfies
\begin{equation}
\tilde a_\mu(t)|0,t>_f =\Lambda (U,t)^{\dagger }a_i(t)\Lambda (U,t)
\Lambda(U,t)^{\dagger }|0>_m=0.
\end{equation}

While Eq.(\ref{eq01x01}) may be viewed as the result of
expanding flavor fields $\phi_\mu(x)$ in the basis parametrized by
free-field mass $m_i$,
it was noticed  that one may as well expand flavor fields in the
basis with the flavor mass parameters 
$m_\mu$ which correspond to choosing 
$u_{\vec k\sigma }^\mu ,v_{-\vec k\sigma }^\mu$ 
as free-field amplitudes with the flavor mass ($m_\mu$) in 
Eqs.(\ref{ladfermion}) and (\ref{ladboson}) \cite{19}. 

In other words, for any $\Lambda (U,t)$,
$\Lambda ^{\prime}(U,t)=I(t)^{-1}\Lambda (U,t)I(t)$, that can be obtained by means of a
similarity transformation mixing $\tilde a_{\mu{\vec k}\sigma}\left( 
t\right) $ and 
$\tilde b_{\mu-\vec k-\sigma}^{\dagger }(t)$ but leaving their combination in $\phi(\vec{k})$ 
unchanged (i.e. $\phi_\mu(\vec k,t)=I(t)^{-1}\phi_\mu(\vec k,t)I(t)$), 
is also a representation of the mixing group. 
The ladder operators, defined by Eqs.(\ref{la01})-(\ref{la04}),
therefore depend on the choice of $I(t)$ or, equivalently, 
the "bare"  mass $m_\mu$ assigned to the flavor fields which is called as a 
mass parametrization.

Although there are different opinions about whether or not the measurable
results of the theory depend on the mass parameters \cite{7,9,19,23}, 
we note that the mass parametrization problem indeed is not specific to
the quantum mixing, but can be revealed in
the free field case as well as in the perturbation theory. 
As discussed in \cite{23}, when dealing with the free field problem 
defined by the free Hamiltonian
\begin{equation}
:H_0:=\sum\limits_{\vec{k}\sigma}\left(\epsilon_{\vec{k}} 
a^{\dagger}_{\vec k\sigma}a_{\vec{k}\sigma} + 
\epsilon_{\vec{k}} b^{\dagger}_{\vec k\sigma}
b_{\vec{k}\sigma}\right),
\end{equation}
one may still consider the change of the mass parametrization
$m\rightarrow m_\mu$ defined in \cite{7} by
\begin{equation}\label{transf}
\left(
\begin{array}{c}
\tilde a(t) \\
\tilde b^{\dagger }(t)
\end{array}
\right) =I^{-1}(t)\left(
\begin{array}{c}
a \\
b^{\dagger }
\end{array}
\right) I(t)=\left(
\begin{array}{cc}
e^{i(\tilde \epsilon_{\vec k}-\epsilon_{\vec k})t}\rho_{\vec k}^{*} & 
  e^{i(\tilde \epsilon_{\vec k}+\epsilon_{\vec k})t}\lambda_{\vec k} \\
e^{-i(\tilde \epsilon_{\vec k}+\epsilon_{\vec k})t}\lambda_{\vec k}^{*} &
  e^{-i(\tilde \epsilon_{\vec k}-\epsilon_{\vec k})t}\rho_{\vec k}  
\end{array}
\right)\left(
\begin{array}{c}
a(0) \\
b^{\dagger }(0)
\end{array}
\right),
\end{equation}
where $\tilde\epsilon_k=\sqrt{k^2+{m_\mu}^2}$ and $ 
\epsilon_k=\sqrt{k^2+m^2}$. 
Indeed, as we observe in \cite{23}, the number operator for the free fields
in this transformation is not conserved, e.g. for fermions
\begin{equation}
<\tilde N>=|\{\tilde a,\tilde a^{\dagger }(t)\}|^2=\left||\rho
_k|^2e^{-i\epsilon _{k}t}+|\lambda _k|^2e^{i\epsilon _{k}t}\right|^2,
\end{equation}
that may lead to obviously wrong conclusion that the number of particles  
in the free field case is not observable quantity.

This can be also understood mathematically, once we note that the above 
transformation is equivalent to the splitting of the initial hamiltonian into
\begin{equation}
H_0=H_{0}^{\prime}+H_I^{\prime}=
\int d^4 p \left( \left\{ (\hat p \psi)^{\dagger}(\hat p \psi)-{m_\mu}^2 
\psi^{\dagger}\psi \right\}+
({m_\mu}^2-m^2) \psi^{\dagger}\psi\right),
\end{equation}
where additional self-interaction term, responsible for oscillation of 
$<\tilde N>$, appears. Physically, the transformation, given by 
Eq.(\ref{transf}), can be viewed as a redefinition of the physical 
one-particle states. The tilde quantities
correspond then to some new quasiparticle objects so that the tilde 
number operator describes a different type of particles and thus it doesn't 
have to be invariant under such transformation. Nevertheless, the charge 
quantum number is still conserved in the transformation, given by 
Eq.(\ref{transf}). The situation here may be analogous to  
the representation of physical observables under the change of coordinate 
systems. Although the Casimir operator (e.g. $\vec S^{2}$ in the spin 
observables) must be independent from the
coordinate system, other physical operators (e.g. $S_x$ $S_y$ and $S_z$)
do depend on the coordinate system. To compare the eigenvalue of $S_z$
between theory and experiment, one should first fix the coordinate system.
Similarly, we think specific mass parameters should be selected 
from the physical reasoning to compare theoretical results ({\it e.g.} the 
occupation number expectation) with experiments. 

From the above example it is clear that the same mass parametrization 
problem is also present in the regular perturbation theory once 
one attempts to redefine the physical one-particle states as shown in 
Eq.(\ref{transf}). Indeed, in the free threory and the perturbation 
theory this issue is resolved by 
the presence of the mass scale of well defined asymptotic physical states, 
which therefore fix the mass
parameters. In the mixing problem, however, at least two feasible mass 
scales may be suggested either by the mass scale of the 
energy-eigenstates or by the flavor mass scale which corresponds 
to no self-interaction term in the hamiltonian, given by 
Eq.(\ref{hamiltonians}), and thus further discussion of this issue in 
the mixing problem is clearly necessary. We think the mass eigenvalues 
that can be measured from the experiments may be the natural choice for 
the mass scale in the given physical system. 


In any case, all the above unified formulation for any number of fields with
integer or half-integer spin holds for the arbitrary mass parameter
$m_\mu $ when 
$\epsilon_i=\sqrt{k^2+m_i^2}$ and $\epsilon_\mu=\sqrt{k^2+m_\mu ^2}$
in Eqs.(\ref{ladshort})-(\ref{ladshort01}) are understood as the energies 
of the free field $\varphi_i$ and the flavor field $\phi_\mu$, respectively.

In the rest of this section, let us consider the explicit form of 
the flavor vacuum state. We
obtain its structure by solving directly the infinite set of equations 
\begin{equation}
\label{vacuumsolve}\tilde a_\nu |0>_f =0,\text{ }\tilde b_\nu |0>_f =0.
\end{equation}
We can express the flavor vacuum state as a linear combination of the
mass eigenstates, i.e. in the most general form,
\begin{equation}\label{eq2_29}
|0>_f =\sum_{\left( n\right) ,(l)}\frac 1{n_1!n_2!\ldots n_k!}B_{\left(
n\right) \left( l\right) }\left( a_1^{\dagger }\right) ^{n_1}\ldots \left(
a_k^{\dagger }\right) ^{n_k}\left( b_{-1}^{\dagger }\right) ^{l_1}\ldots
\left( b_{-k}^{\dagger }\right) ^{l_k}|0>_m,
\end{equation}
with $(n)=(n_1n_2n_3\ldots )$. After applying Eq.(\ref{vacuumsolve}) to 
Eq.(\ref{eq2_29}) we get an infinite set of equations given by
\begin{equation}
\sum_j(\alpha _{\mu j}B_{(n_j+1)(l)}+\beta _{\mu j}B_{(n)(l_j-1)})=0,
\text{ for all sets of }(n),(l),
\end{equation}
where $(n_j\pm 1)=(n_1n_2\ldots n_j\pm 1\ldots)$. The solution of this problem 
is presented
in the Appendix B. For the flavor vacuum state we find 
\begin{equation}
|0>_f =\frac 1{{\cal Z}}\exp (\sum_{i,j=1}^NZ_{ij}a_i^{\dagger
}b_{-j}^{\dagger })|0>_m ,
\end{equation}
where $Z_{ij}$ is an $(i,j)$ element of the matrix $\hat Z=-\hat \alpha ^{-1}\hat \beta$.
The normalization constant ${\cal Z}$ is fixed by 
$_f <0|0>_f =1$; ${\cal Z}=\det ^{1/2}\left( 1+\hat Z
\hat Z^{\dagger }\right) $ for fermions and 
${\cal Z}=\det ^{-1/2}\left( 1-\hat Z\hat Z^{\dagger }\right) $ for bosons. 
The flavor Fock-space is then built
by applying the flavor-field creation operators ($\tilde a_\mu ^{\dagger }
$,$\tilde b_\nu ^{\dagger }$) to the vacuum state $|0>_f$.

We see that the flavor vacuum state has a rich coherent structure. This
situation is different from the perturbative quantum field theory, where 
the adiabatic enabling of interaction is present and
$|0>_{interacting}\sim |0>_{free}$. The nonperturbative
vacuum solution renders non-trivial effects in the flavor 
dynamics as we will show in Section III. In particular, the
normalization constant ${\cal Z}$ is always greater than 1 so that in the
infinite volume limit, when density of states is going to infinity, we have 
\begin{equation}
{\cal Z}_{tot}=\exp \left( \frac V{(2\pi )^3}\int d\vec k\ln ({\cal Z}_{\vec 
k})\right) \rightarrow \infty.
\end{equation}
Thus, any possible state for the flavor vacuum shall have infinite norm
in the free-field Fock space and therefore the flavor vacuum 
state cannot be found in original Fock
space. The unitary inequivalence of the flavor Fock space and the original
Fock space is therefore established, i.e. $_f<0|0>_m=\frac{1}{{\cal 
Z}_{tot}}\rightarrow 0$. The effect is essentially due to an
infinite number of momentum degrees of freedom, which is analogous to 
the existence of phase transition in the infinite volume limit.

\setcounter{equation}{0} \setcounter{figure}{0} 
\renewcommand{\theequation}{\mbox{3.\arabic{equation}}} 
\renewcommand{\thefigure}{\mbox{3.\arabic{figure}}}

\section{Time Dynamics of the Mixed Quantum Fields}

Now we have a closer look at the dynamics of quantum 
fields respresented by the ladder operators shown in 
Eq.($\ref{ladshort}$). First of all, we note 
that only $a_{i\vec k\sigma}$ and $b_{i-\vec k-\sigma}$ operators and 
their conjugates are mixed together. We
denote the set of quantum fields formed by all linear combinations of these
operators and their products (algebra on $a_{i\vec k\sigma}$, 
$b_{i-\vec k-\sigma}$ and h.c.) as
a cluster $\Omega_{\vec k\sigma}$ with a particular momentum 
$\vec k$ and a particular helicity $\sigma$. It
follows that $\Omega _{\vec k\sigma }$'s are invariant under
mixing transformation $\Lambda (U,t)$ and we thus can treat each cluster
independently from each other.

The time dynamics of the flavor fields is determined by the non-equal time
commutation/anticommutation relationships for boson/fermion fields that can be derived 
from Eq.($\ref{ladshort}$) using the standard commutation/anticommutation relationships 
for the original ladder operators;
\begin{equation}
\label{commutators}
\begin{array}{c}
F_{\mu \nu }(t)=
[\tilde a_\mu(t) ,\tilde a_\nu ^{\dagger }]_{\pm}=
\sum\limits_{k,k^{\prime }}\left( \alpha _{\mu k}\alpha _{\nu k^{\prime
}}^{*}\left[ a_k e^{-i\epsilon _{k}t},a_{k^{\prime }}^{\dagger }\right] _{\pm }+
\beta _{\mu k}\beta _{\nu k^{\prime }}^{*}
\left[b_{-k}^{\dagger }e^{i\epsilon _{k}t},b_{-k^{\prime }}\right] _{\pm
}\right)  \\ =\sum\limits_k(\alpha _{\mu k}\alpha _{\nu k}^{*}e^{-i\epsilon
_kt}-\left( -1\right) ^{2S}\beta _{\mu k}\beta _{\nu k}^{*}e^{i\epsilon
_kt}); \\ 
\lbrack 
\tilde b_{-\mu }(t),\tilde b_{-\nu }^{\dagger }]_{\pm }=F_{\nu
\mu }(t); \\ 
G_{\mu \nu }(t)=[
\tilde b_{-\mu }(t),\tilde a_\nu ]_{\pm
}=\sum\limits_{k,k^{\prime }}\left( \alpha _{\mu k}^{*}\beta _{\nu k^{\prime
}}\left[ b_{-k}e^{-i\epsilon _{k}t},b_{-k^{\prime }}^{\dagger }\right] _{\pm }+
\left( -1\right) ^{2S}\beta _{\mu k}^{*}\alpha _{\nu
k^{\prime }}\left[ a_k^{\dagger }e^{i\epsilon _{k}t},a_{k^{\prime }}\right] _{\pm }\right)  
\\ =\sum\limits_k(\alpha _{\mu k}^{*}\beta _{\nu
k}e^{-i\epsilon _kt}-\beta _{\mu k}^{*}\alpha _{\nu k}e^{i\epsilon _kt}).
\end{array}
\end{equation}
The two matrices $\hat F$ and $\hat G$ represent the only nontrivial
commutators/anticommutators in the sense that all others are either zero or
can be written in terms of the elements of these matrices. It is useful to
note that, for $t=0$, Eq.(\ref{commutators}) shall be reduced to $F_{\mu \nu
}(0)=\delta _{\mu \nu }$ and $G_{\mu \nu }(0)=0$. We also note that
\begin{equation}
\begin{array}{c}
F_{\mu \nu }(t)^{*}=F_{\nu \mu }(-t), \\ 
G_{\mu \nu }(t)^{*}=-G_{\nu \mu }(t).
\end{array}
\end{equation}

Eq.(\ref{commutators}) allows us to compute many mixing quantities directly. 
The time dynamics of the flavor-field ladder operators can be derived by
writing them as $\tilde a_\mu \left( t\right) =\sum\limits_\nu (f_{\mu \nu }
\tilde a_\nu (0)+g_{\mu \nu }\tilde b_{-\nu }^{\dagger }(0)+\ldots )$. Then,
one can get
straightforwardly $f_{\mu \nu }^{*}=[\tilde a_\nu (0),
\tilde a_\mu ^{\dagger}\left( t\right) ]_{\pm }=F_{\nu \mu }(-t)$
 and 
 $g_{\mu \nu }=[\tilde b_{-\nu }(0),\tilde a_\mu \left( t\right) ]_{\pm }=
 G_{\nu \mu }(-t)$ while all other coefficients are zeros: 
\begin{equation}
\label{ladself}
\begin{array}{c}
\tilde a_\mu \left( t\right) =\sum\limits_\nu \left( F_{\mu\nu }(
t)\tilde a_\nu +G_{\nu \mu }(-t) \tilde b_{-\nu
}^{\dagger }\right) ; \\ 
\tilde b_{-\mu }\left( t\right) =\sum\limits_\nu
\left( F_{\nu\mu}( t)\tilde b_{-\nu }+\left( -1\right)
^{2S}G_{\mu \nu }(t) \tilde a_{\nu}^{\dagger }\right) .
\end{array}
\end{equation}

We now consider the
condensate densities of the definite-mass particles in the flavor vacuum ($
Z_i^{\prime }=_f <0|a_i^{\dagger }\left( t\right) a_i\left( t\right)
|0>_f $), the number of definite-flavor particles in the flavor vacuum 
($Z_\nu =_f <0|\tilde a_\nu ^{\dagger }\left( t\right) \tilde a_\nu \left(
t\right) |0>_f $) and the particle number average for a single definite-flavor 
particle initial state, which is related in the Heisenberg picture to 
$N_{\rho \nu \sigma }=_\mu <\rho |\tilde a_\nu ^{\dagger }\left( t\right) 
\tilde a_\nu \left( t\right) |\sigma >_\mu $, $\bar N_{\rho \nu \sigma
}=_\mu <\rho |\tilde b_{-\nu }^{\dagger }\left( t\right) \tilde b_{-\nu
}\left( t\right) |\sigma >_\mu $. 

The free-field particle condensates in the flavor vacuum state 
are computed from the explicit
form of the ladder operators given by Eq.(\ref{ladshort}) as 
\begin{equation}\label{eq3_3}
Z_i^{\prime }=\sum_j\left| \beta _{ij}\right| ^2.
\end{equation}
In the following, the particle-antiparticle
symmetry should be accounted for, so that 
a corresponding antiparticle quantity can be found
from the particle expression after a
necessary substitution (particles$\rightarrow $antiparticles and vice
versa).
Thus, the antiparticle condensate is given by the same quantity in 
Eq.(\ref{eq3_3}). The definite-flavor particle condensates in the 
free-field vacuum is also given by Eq.(\ref{eq3_3}). 

Using Eq.(\ref{ladself}), we get the flavor-field condensates in the 
flavor vacuum ($Z_{\nu}$) as 
\begin{equation}
Z_\nu (t)=\sum_\mu \left| G_{\nu \mu }(-t)\right| ^2.
\end{equation}
It is remarkable that this number is not zero but oscillates, demonstrating
the oscillations of definite-flavor particles in the flavor vacuum. This effect
reveals the unitary inequivalence of the flavor Fock-spaces for different
times due to the time-dynamics of the flavor vacuum.

The evolution of the particle ($N_{\rho \nu \sigma }$) and
antiparticle ($\bar N_{\rho \nu \sigma }$) number with flavor $\nu $ can be found
using the standard technique of normal ordering, i.e. moving annihilation operators 
to the right side
and creation operators to the left side of the expression. 
With this technique, we obtain
\begin{equation}
\begin{array}{c}
\begin{array}{c}
N_{\rho \nu \sigma }(t)=[
\tilde a_\rho ,\tilde a_\nu ^{\dagger }\left( t\right) ]_{\pm }[\tilde a_\nu
\left( t\right) ,\tilde a_\sigma ^{\dagger }]_{\pm }+\delta _{\rho \sigma
}<0|\tilde a_\nu ^{\dagger }\left( t\right) \tilde a_\nu \left( t\right)
|0>= \\ 
=F_{\nu\rho}^{*}(t)F_{\nu\sigma}(t)+\delta _{\rho \sigma
}Z_\nu (t),
\end{array}
\\ 
\begin{array}{c}
\bar N_{\rho \nu \sigma }(t)=\left( -1\right) ^{2S}[\tilde a_\rho ,\tilde b
_{-\nu }\left( t\right) ]_{\pm }[\tilde b_{-\nu }^{\dagger }\left( t\right) ,
\tilde a_\sigma ^{\dagger }]_{\pm }+\delta _{\rho \sigma }<0|\tilde b_\nu
^{\dagger }\left( t\right) \tilde b_\nu \left( t\right) |0>= \\ 
=\left(-1\right) ^{2S}G_{\nu \rho }(t)G_{\nu \sigma }(t)^{*}+
\delta _{\rho \sigma}Z_\nu (t).
\end{array}
\end{array}
\end{equation}
The flavor charge $Q_{\rho\nu\sigma}=N_{\rho\nu\sigma}-{\bar 
N}_{\rho\nu\sigma}$ \cite{7,8,9} is then given by: 
\begin{equation}
Q_{\rho \nu \sigma }=N_{\rho \nu \sigma }-\bar N_{\rho \nu \sigma }=
F_{\nu\rho}^{*}(t)F_{\nu\sigma}(t)-\left( -1\right) ^{2S}G_{\nu \rho}(t)
G_{\nu \sigma }(t)^{*}.
\end{equation}
For a specific case of the number evolution in the beam with
a fixed 3-momentum, we find: 
\begin{equation}
\label{XX01}
\begin{array}{c}
N_{\rho \nu \rho }=<0|
\tilde a_\rho \tilde a_\nu ^{\dagger }\left( t\right) \tilde a_\nu \left(
t\right) \tilde a_\rho ^{\dagger }|0>=
\left| F_{\nu\rho}(t)\right|^2+Z_\nu (t), \\ 
\bar N_{\rho \nu \rho }=<0|\tilde a_\rho \tilde b_{-\nu
}^{\dagger }\left( t\right) \tilde b_{-\nu }\left( t\right) \tilde a_\rho
^{\dagger }|0>=\left( -1\right) ^{2S}\left| G_{\nu \rho }(t)\right|
^2+Z_\nu (t), \\ 
Q_{\rho \nu \rho }=\left| F_{\nu\rho}(t)\right|^2-
\left( -1\right) ^{2S}\left| G_{\nu \rho }(t)\right| ^2.
\end{array}
\end{equation}
We note that $N_{\rho \nu \rho }$'s as well as 
$Q_{\rho \nu \rho }$'s are in general dependent on the choice of mass
parameter $m_{\mu}$.

We may explicitly see this in the example of the charge operator. 
According to Eq.(\ref{XX01}), we get 
\begin{equation}
\begin{array}{l}
Q_{\mu \nu \mu }=\sum\limits_{k,k^{\prime }}(\alpha _{\mu k}\alpha _{\nu
k}^{*}e^{i\epsilon _kt}-\left( -1\right) ^{2S}\beta _{\mu k}\beta _{\nu
k}^{*}e^{-i\epsilon _kt})(\alpha _{\mu k^{\prime }}^{*}\alpha _{\nu k^{\prime
}}e^{-i\epsilon _{k^{\prime }}t}-\left( -1\right) ^{2S}\beta _{\mu k^{\prime
}}^{*}\beta _{\nu k^{\prime }}e^{i\epsilon _{k^{\prime }}t})- \\ 
-(-1)^{2S}\sum\limits_{k,k^{\prime }}(\alpha _{\nu k}^{*}\beta _{\mu
k}e^{-i\epsilon _kt}-\beta _{\nu k}^{*}\alpha _{\mu k}e^{i\epsilon
_kt})(\alpha _{\nu k^{\prime }}\beta _{\mu k^{\prime }}^{*}e^{i\epsilon
_{k^{\prime }}t}-\beta _{\nu k^{\prime }}\alpha _{\mu k^{\prime
}}^{*}e^{-i\epsilon _{k^{\prime }}t})= \\ 
\begin{array}{r}
=\sum\limits_{k,k^{\prime }}e^{-i(\epsilon _{k^{\prime }}-\epsilon
_k)t}(\alpha _{\mu k^{\prime }}^{*}\alpha _{\nu k^{\prime }}\alpha _{\mu
k}\alpha _{\nu k}^{*}-(-1)^{2S}\beta _{\nu k^{\prime }}\alpha _{\mu
k^{\prime }}^{*}\beta _{\nu k}^{*}\alpha _{\mu k})+ \\ 
e^{i(\epsilon _{k^{\prime }}-\epsilon _k)t}(\beta _{\mu k}\beta _{\nu
k}^{*}\beta _{\mu k^{\prime }}^{*}\beta _{\nu k^{\prime }}-(-1)^{2S}\alpha
_{\nu k^{\prime }}\beta _{\mu k^{\prime }}^{*}\alpha _{\nu k}^{*}\beta _{\mu
k})- \\ 
\left( -1\right) ^{2S}e^{-i(\epsilon _{k^{\prime }}+\epsilon _k)t}(\beta
_{\mu k}\beta _{\nu k}^{*}\alpha _{\mu k^{\prime }}^{*}\alpha _{\nu
k^{\prime }}-\alpha _{\nu k}^{*}\beta _{\mu k}\beta _{\nu k^{\prime }}\alpha
_{\mu k^{\prime }}^{*})- \\ 
\left( -1\right) ^{2S}e^{i(\epsilon _{k^{\prime }}+\epsilon _k)t}(\alpha
_{\mu k}\alpha _{\nu k}^{*}\beta _{\mu k^{\prime }}^{*}\beta _{\nu k^{\prime
}}-\beta _{\nu k}^{*}\alpha _{\mu k}\alpha _{\nu k^{\prime }}\beta _{\mu
k^{\prime }}^{*})
\end{array}
\\ 
\begin{array}{r}
=\sum\limits_{k,k^{\prime }}e^{-i(\epsilon _{k^{\prime }}-\epsilon
_k)t}\alpha _{\mu k^{\prime }}^{*}\alpha _{\mu k}(\alpha _{\nu k^{\prime
}}\alpha _{\nu k}^{*}-(-1)^{2S}\beta _{\nu k^{\prime }}\beta _{\nu k}^{*})-
\\ 
(-1)^{2S}e^{i(\epsilon _{k^{\prime }}-\epsilon _k)t}\beta _{\mu k}\beta
_{\mu k^{\prime }}^{*}(\alpha _{\nu k^{\prime }}\alpha _{\nu
k}^{*}-(-1)^{2S}\beta _{\nu k}^{*}\beta _{\nu k^{\prime }})- \\ 
\left( -1\right) ^{2S}e^{-i(\epsilon _{k^{\prime }}+\epsilon _k)t}\beta _{\mu
k}\alpha _{\mu k^{\prime }}^{*}(\beta _{\nu k}^{*}\alpha _{\nu k^{\prime
}}-\alpha _{\nu k}^{*}\beta _{\nu k^{\prime }})- \\ 
\left( -1\right) ^{2S}e^{i(\epsilon _{k^{\prime }}+\epsilon _k)t}\alpha
_{\mu k}\beta _{\mu k^{\prime }}^{*}(\alpha _{\nu k}^{*}\beta _{\nu
k^{\prime }}-\beta _{\nu k}^{*}\alpha _{\nu k^{\prime }})
\end{array}
\\ 
\begin{array}{r}
=\sum\limits_{k,k^{\prime }}(\alpha _{\nu k^{\prime }}\alpha _{\nu
k}^{*}-(-1)^{2S}\beta _{\nu k^{\prime }}\beta _{\nu k}^{*})(e^{-i(\epsilon
_{k^{\prime }}-\epsilon _k)t}\alpha _{\mu k^{\prime }}^{*}\alpha _{\mu
k}-(-1)^{2S}e^{i(\epsilon _{k^{\prime }}-\epsilon _k)t}\beta _{\mu k}\beta
_{\mu k^{\prime }}^{*})- \\ 
\left( -1\right) ^{2S}(\beta _{\nu k}^{*}\alpha _{\nu k^{\prime }}-\alpha
_{\nu k}^{*}\beta _{\nu k^{\prime }})(e^{-i(\epsilon _{k^{\prime }}+\epsilon
_k)t}\beta _{\mu k}\alpha _{\mu k^{\prime }}^{*}-e^{i(\epsilon _{k^{\prime
}}+\epsilon _k)t}\alpha _{\mu k}\beta _{\mu k^{\prime }}^{*}).
\end{array}
\end{array}
\end{equation}
Taking into account Eq.(\ref{eq03x01}), we can write, e.g. for
fermions (S=1/2) 
$$
\begin{array}{c}
\alpha _{\nu k^{\prime }}\alpha _{\nu k}^{*}+\beta _{\nu k^{\prime }}\beta
_{\nu k}^{*}=U_{\nu k^{\prime }}U_{\nu k}^{*}(\cos (\theta _{\nu k^{\prime
}})\cos (\theta _{\nu k})+\sin (\theta _{\nu k^{\prime }})\sin (\theta _{\nu
k}))= \\ 
=U_{\nu k^{\prime }}U_{\nu k}^{*}\cos (\theta _{\nu k^{\prime }}-\theta
_{\nu k})=U_{\nu k^{\prime }}U_{\nu k}^{*}\cos (\theta _{kk^{\prime }}), \\ 
\beta _{\nu k}^{*}\alpha _{\nu k^{\prime }}-\alpha _{\nu k}^{*}\beta _{\nu
k^{\prime }}=U_{\nu k^{\prime }}U_{\nu k}^{*}(\cos (\theta _{\nu k^{\prime
}})\sin (\theta _{\nu k})-\cos (\theta _{\nu k^{\prime }})\sin (\theta _{\nu
k}))= \\ 
=U_{\nu k^{\prime }}U_{\nu k}^{*}\sin (\theta _{\nu k}-\theta _{\nu
k^{\prime }})=U_{\nu k^{\prime }}U_{\nu k}^{*}\sin (\theta _{k^{\prime }k}).
\end{array}
$$
Thus, we find
\begin{equation}
\begin{array}{r}
Q_{\mu \nu \mu }=\sum\limits_{k,k^{\prime }}U_{\nu k^{\prime }}U_{\nu
k}^{*}U_{\mu k}U_{\mu k^{\prime }}^{*}(\cos ^2(\theta _{kk^{\prime }})\cos
(\omega _{k^{\prime }k}t)+i\cos (\theta _{k^{\prime }k})\cos (\theta _{\mu
k}+\theta _{\mu k^{\prime }})\sin (\omega _{kk^{\prime }}t)+ \\ 
\sin ^2(\theta _{k^{\prime }k})\cos (\Omega _{k^{\prime }k}t)-i\sin (\theta
_{k^{\prime }k})\sin (\theta _{\mu k}+\theta _{\mu k^{\prime }})\sin (\Omega
_{kk^{\prime }}t));
\end{array}
\end{equation}
where $\Omega _{ij}=\epsilon _i+\epsilon _j$ and $\omega _{ij}=\epsilon
_i-\epsilon _j$. This can be rewritten as 
\begin{equation}
\begin{array}{l}
Q_{\mu \nu \mu }=\sum\limits_{k,k^{\prime }}Re(U_{\nu k^{\prime }}U_{\nu
k}^{*}U_{\mu k}U_{\mu k^{\prime }}^{*})(\cos ^2(\theta _{kk^{\prime }})\cos
(\omega _{k^{\prime }k}t)-(-1)^{2S}\sin ^2(\theta _{k^{\prime }k})\cos
(\Omega _{k^{\prime }k}t))+ \\ 
+\sum\limits_{k,k^{\prime }}Im(U_{\nu k^{\prime }}U_{\nu k}^{*}U_{\mu
k}U_{\mu k^{\prime }}^{*})(\cos (\theta _{kk^{\prime }})\cos (\theta _{\mu
k}+\theta _{\mu k^{\prime }})\sin (\omega _{k^{\prime }k}t)-(-1)^{2S}\sin
(\theta _{k^{\prime }k})\sin (\theta _{\mu k}+\theta _{\mu k^{\prime }})\sin
(\Omega _{k^{\prime }k}t)).
\end{array}
\end{equation}
This formula is also valid for bosons with the substitution of $\cos
\rightarrow \cosh $, $\sin \rightarrow \sinh$.

We see now that $Q_{\mu \nu \mu }$ does not depend on the mass parameters
only for real mixing matrices $U_{\mu k}$ \cite{8,19}. Otherwise,
there is a nontrivial mass dependence from the imaginary part of $U$. 
Interestingly, even in the latter case, there is no 
dependence on the mass of the flavor field $\nu $ ($m_\nu $) but only 
on the mass of the initial flavor state $\mu $. 

We also note that Eq.(\ref{XX01}) may be viewed as a superposition of
the two terms: $\rho\rightarrow \nu$ propagation and background vacuum contribution 
$Z_{\nu}$. Thus, one may introduce the particle-particle and particle-antiparticle 
propagation amplitudes, respectively, defined by 
\begin{equation}
\begin{array}{c}
{\cal P}_{\rho \rightarrow \nu }(k,t)=[ 
\tilde a_\nu (t),\tilde a_\rho ^{\dagger }(0)]_{\pm }=F_{\nu \rho }(t)
\\ 
{\cal P}_{\rho \rightarrow -\bar{\nu} }(k,t)=
[\tilde b_{-\nu} (t),\tilde a_\rho (0)]_{\pm }=G_{\nu \rho }(t). 
\end{array}
\end{equation}
Indeed, such propagation amplitudes appear from the flavor-field Green
function $_f <0(t=0)|\phi _\nu (k,t)\phi _\rho ^{\dagger
}(k,0)|0(t=0)>_f$ for $t>0$. 
Propagation functions, defined in this way, are clearly the Green functions
of the mixed-field problem and obey the causality features relevant to such
Green functions
\footnote{See Refs.\cite{5,19} for the discussion of the Green functions in 
the quantum theory of the mixed fields.}.

\setcounter{equation}{0} \setcounter{figure}{0} 
\renewcommand{\theequation}{\mbox{4.\arabic{equation}}} 
\renewcommand{\thefigure}{\mbox{4.\arabic{figure}}}

\section{Two-field unitary mixing}

\subsection{Vector Meson Mixing (S=1)}

We now consider the unitary mixing of 2 fields with spin 1 (vector mesons). U(2)
parametrization consists of 4 parameters: 3 phases that can be absorbed in the
phase redefinition of fields and one essential real angle that is left, so that
\begin{equation}
U=\left( 
\begin{array}{cc}
\cos (\theta ) & \sin (\theta ) \\ 
-\sin (\theta ) & \cos (\theta )
\end{array}
\right) . 
\end{equation}
Using Appendix A, we then define 
$\gamma _{\mu i}^{\pm }=\frac 12\left(\sqrt{\frac{\epsilon _\mu }{\epsilon _i}}
\pm \sqrt{\frac{\epsilon _i}{\epsilon _\mu }}\right) $ for $\sigma =\pm 1$ and 
\begin{equation}
\begin{array}{c}
\gamma _{\mu i}^{+}=
\frac 12\frac{\epsilon _\mu \epsilon _i-k^2}{m_\mu m_i}\left( \sqrt{\frac{
\epsilon _\mu}{\epsilon _i}}+\sqrt{\frac{\epsilon _i}{\epsilon 
_\mu}}\right) ,
\\ \gamma _{\mu i}^{-}=\frac 12\frac{k^2+\epsilon _\mu \epsilon _i}{m_\mu 
m_i}\left( \sqrt{\frac{\epsilon _\mu}{\epsilon _i}}-\sqrt{\frac{\epsilon 
_i}{ \epsilon _\mu}}\right) 
\end{array}
\end{equation}
for $\sigma =0$. For the free field mass $m_i$ basis, 
$\gamma _{12}^{+}=\gamma _{21}^{+}=\gamma _{+}$, $\gamma _{12}^{-}=-\gamma
_{21}^{-}=\gamma _{-}$. 
We use this basis in Section IV. 

The ladder mixing matrices $\alpha $ and $\beta $ are given by
\begin{equation}
\begin{array}{c}
\alpha =\left( 
\begin{array}{cc}
\cos \left( \theta \right)  & \gamma _{+}\sin \left( \theta \right) 
\\ -\gamma _{+}\sin \left( \theta \right)  & \cos \left( \theta
\right) 
\end{array}
\right) , \\ 
\beta =\left( 
\begin{array}{cc}
0 & \gamma _{-}\sin \left( \theta \right)  \\ -\gamma
_{-}\sin \left( \theta \right)  & 0
\end{array}
\right) .
\end{array}
\end{equation}
For the flavor charge oscillation, we then obtain the result that is not 
dependent on the mass parametrization:
\begin{equation}\label{eq4_2}
\begin{array}{c}
Q_{111}=1+
\sin ^2\left( 2\theta \right) \left( \gamma _{-}^2\sin ^2\left( 
\frac{\Omega _{12}t}2\right) -\gamma _{+}^2\sin ^2\left( \frac{\omega _{12}t}
2\right) \right) , \\ 
Q_{121}=\sin ^2\left( 2\theta \right) \left(
\gamma _{+}^2\sin ^2\left( \frac{\omega _{12}t}2\right) -\gamma _{-}^2\sin
^2\left( \frac{\Omega _{12}t}2\right) \right) .
\end{array}
\end{equation}
We see that this result, with an exception of greater complexity of $\gamma
_{\pm }$, is identical to the case of spin 0 \cite{7,23}. According to the above
theory, in fact, this should be the case for the two-field mixing with 
any integer spin. For $S=1$ we see that an essential difference from the 
scalar/pseudoscalar meson mixing, such as the complication of momentum 
dependence of 
$\gamma _{\pm }$, occurs only for the mixing of longitudinally polarized 
particles. The mixing of transverse
components is essentially same as in the case of spin-zero particles.

The details of non-equal time commutators are given by 
\begin{equation}\label{eq4_3}
F=\left\{ 
\begin{array}{c}
\begin{array}{cc}
e^{-i\epsilon _1t}\cos ^2\left( \theta \right) +e^{-i\epsilon _2t}
\gamma _{+}^2\sin ^2\left( \theta \right) -e^{i\epsilon _2t}
\gamma_{-}^2\sin ^2\left( \theta \right) ; & \gamma _{+}\sin \left(
\theta \right) \cos \left( \theta \right) \left( e^{-i\epsilon
_2t}-e^{-i\epsilon _1t}\right) 
\end{array}
\\ 
\begin{array}{cc}
\gamma _{+}\sin \left( \theta \right) \cos \left( \theta \right)
\left( e^{-i\epsilon _2t}-e^{-i\epsilon _1t}\right) ; & e^{-i\epsilon
_2t}\cos ^2\left( \theta \right) +e^{-i\epsilon _1t}\gamma _{+}^2
\sin ^2\left( \theta \right) -e^{i\epsilon _1t}\gamma _{-}^2\sin
^2\left( \theta \right) 
\end{array}
\end{array}
\right\} ,
\end{equation}\label{eqCDE}
\begin{equation}
G=\left( 
\begin{array}{cc}
\gamma _{+}\gamma _{-}\sin ^2\left( \theta \right) \left(
e^{-i\epsilon _2t}-e^{i\epsilon _2t}\right)  & \gamma _{-}\sin
\left( \theta \right) \cos \left( \theta \right) \left( e^{-i\epsilon
_1t}-e^{i\epsilon _2t}\right)  \\ 
\gamma _{-}\sin \left( \theta
\right) \cos \left( \theta \right) \left( e^{-i\epsilon _2t}-e^{i\epsilon
_1t}\right)  & \gamma _{+}\gamma _{-}\sin ^2\left( \theta \right)
\left( e^{i\epsilon _1t}-e^{-i\epsilon _1t}\right) 
\end{array}
\right) .
\end{equation}
The condensates of free-field particles are 
\begin{equation}
Z_1^{\prime }=Z_2^{\prime }=\gamma _{-}^2\sin ^2\left( \theta
\right) 
\end{equation}
and the condensates of the flavor particles in the vacuum are 
\begin{equation}
\begin{array}{c}
Z_1=4\gamma _{-}^2\sin ^2\left( \theta \right) \left( \cos ^2\left( \theta
\right) \sin ^2\left( 
\frac{\Omega _{12}t}2\right) +\gamma _{+}^2\sin ^2\left( \theta
\right) \sin ^2\left( \frac{\Omega _{22}t}2\right) \right) , \\ 
Z_2=4\gamma
_{-}^2\sin ^2\left( \theta \right) \left( \cos ^2\left( \theta \right) \sin
^2\left( \frac{\Omega _{12}t}2\right) +\gamma _{+}^2\sin ^2\left(
\theta \right) \sin ^2\left( \frac{\Omega _{11}t}2\right) \right) .
\end{array}
\end{equation}
The flavor vacuum structure is defined by the matrix $\hat Z$: 
\begin{equation}\label{eq4_7}
\hat Z=\frac {-1}{\left( \cos ^2\left( \theta \right) +\gamma _{+}^2
\sin ^2\left( \theta \right) \right) }\left( 
\begin{array}{cc}
-\gamma _{+}\gamma _{-}\sin ^2\left( \theta \right)  & \gamma
_{-}\cos \left( \theta \right) \sin \left( \theta \right)  \\ 
\gamma _{-}\cos \left( \theta \right) \sin \left( \theta \right)  & 
\gamma _{+}\gamma _{-}\sin ^2\left( \theta \right) 
\end{array}
\right) 
\end{equation}
with the normalization constant being ${\cal Z}=(1-\frac{\gamma _{-}^2\sin
^2(\theta )}{\cos ^2(\theta )+\gamma _{+}^2\sin ^2(\theta )})^{-1}=1+\gamma
_{-}^2\sin ^2(\theta )$.

The time evolution of the flavor particle number (if \#1 was emitted) is given
by: 
\begin{equation}\label{eq4_8}
\begin{array}{c}
N_{111}=1+\sin ^2\left( \theta \right) \{8\gamma _{-}^2\cos ^2\left( \theta
\right) \sin ^2\left( 
\frac{\Omega _{12}t}2\right) -4\gamma _{+}^2\cos ^2\left( \theta \right) \sin
^2\left( \frac{\omega _{12}t}2\right) + \\ +
8\gamma _{+}^2\gamma _{-}^2\sin ^2\left( \theta \right) \sin ^2\left( 
\frac{\Omega _{22}t}2\right) \}, 
\\ \bar N_{111}=4\gamma _{-}^2\sin ^2\left(
\theta \right) \left( 2\gamma _{+}^2\sin ^2\left( \theta \right) \sin
^2\left( \frac{\Omega _{22}t}2\right) +\cos ^2\left( \theta \right) \sin
^2\left( \frac{\Omega _{12}t}2\right) \right) ,
\end{array}
\end{equation}\label{eqABC}
\begin{equation}
\begin{array}{c}
N_{121}=\sin ^2\left( \theta \right) \{4\gamma _{+}^2\cos ^2\left( \theta
\right) \sin ^2\left( 
\frac{\omega _{12}t}2\right) +4\gamma _{-}^2\cos ^2\left( \theta \right) \sin
^2\left( \frac{\Omega _{12}t}2\right) + \\ +
4\gamma _{+}^2\gamma _{-}^2\sin ^2\left( \theta \right) \sin ^2\left( 
\frac{\Omega _{11}t}2\right) , \\
 \bar N_{121}=4\gamma _{-}^2\sin ^2\left(
\theta \right) \left( 2\cos ^2\left( \theta \right) \cos ^2\left( \frac{
\Omega _{12}t}2\right) +\gamma _{+}^2\sin ^2\left( \theta \right)
\sin ^2\left( \frac{\Omega _{11}t}2\right) \right) .
\end{array}
\end{equation}

Also we note that the scalar and pseudoscalar case follows immediately  
from the above
presentation when $\gamma _{\mu i}^{\pm }=\frac 12\left( \sqrt{\frac{\epsilon
_\mu }{\epsilon _i}}\pm \sqrt{\frac{\epsilon _i}{\epsilon _\mu }}\right) $.
In this respect, the spin-zero mixing is equivalent to the mixing of 
transverse components 
of vector fields, described by Eqs.(\ref{eq4_2}), (\ref{eq4_3}), 
(\ref{eq4_7}) and (\ref{eq4_8}).
These results are consistent with the previously known results \cite{7,23}.

\subsection{Fermion mixings (S=1/2)}

We also present here the calculations for $s=1/2$ case. For the consistent 
notation with the previous works \cite{4,6}
\footnote{In our notation, $U=\gamma _{+}$, $V=\gamma _{-}$}, we define  
\begin{equation}
\begin{array}{c}
U=
\frac{\sqrt{\left( \epsilon _1+m_1\right) \left( \epsilon _2+m_2\right) }+
\sqrt{\left( \epsilon _1-m_1\right) \left( \epsilon _2-m_2\right) }}{2\sqrt{
\epsilon _1\epsilon _2}}, \\ V=\sigma \frac{\sqrt{\left( \epsilon
_1-m_1\right) \left( \epsilon _2+m_2\right) }-\sqrt{\left( \epsilon
_1+m_1\right) \left( \epsilon _2-m_2\right) }}{2\sqrt{\epsilon _1\epsilon _2}
}.
\end{array}
\end{equation}
The charge fluctuations are then given by 
\begin{equation}
\begin{array}{c}
Q_{111}=1-\sin ^2\left( 2\theta \right) \left( U^2\sin ^2\left( 
\frac{\omega _{12}t}2\right) +V^2\sin ^2\left( \frac{\Omega _{12}t}2\right)
\right) , \\ Q_{121}=\sin ^2\left( 2\theta \right) \left( U^2\sin ^2\left( 
\frac{\omega _{12}t}2\right) +V^2\sin ^2\left( \frac{\Omega _{12}t}2\right)
\right) 
\end{array}
\end{equation}
and the ladder mixing matrices are
\begin{equation}
\begin{array}{c}
\alpha =\left( 
\begin{array}{cc}
\cos \left( \theta \right)  & U\sin \left( \theta \right)  \\ 
-U\sin \left( \theta \right)  & \cos \left( \theta \right) 
\end{array}
\right) , \\ 
\beta =\left( 
\begin{array}{cc}
0 & V\sin \left( \theta \right)  \\ 
V\sin \left( \theta \right)  & 0
\end{array}
\right) ,
\end{array}
\end{equation}
which are same with the previously known results \cite{4,6}.

We can also give more details on the fermion mixing dynamics. The non-equal 
time anticommutators are given by 
\begin{equation}
F=\left\{ 
\begin{array}{c}
\begin{array}{cc}
e^{-i\epsilon _1t}\cos ^2\left( \theta \right) +e^{-i\epsilon _2t}U^2\sin
^2\left( \theta \right) +e^{i\epsilon _2t}V^2\sin ^2\left( \theta \right) ;
& U\sin \left( \theta \right) \cos \left( \theta \right) \left(
e^{-i\epsilon _2t}-e^{-i\epsilon _1t}\right) 
\end{array}
\\ 
\begin{array}{cc}
U\sin \left( \theta \right) \cos \left( \theta \right) \left( e^{-i\epsilon
_2t}-e^{-i\epsilon _1t}\right) ; & e^{-i\epsilon _2t}\cos ^2\left( \theta
\right) +e^{-i\epsilon _1t}U^2\sin ^2\left( \theta \right) +e^{i\epsilon
_1t}V^2\sin ^2\left( \theta \right) 
\end{array}
\end{array}
\right\} ,
\end{equation}
\begin{equation}
G=\left( 
\begin{array}{cc}
UV\sin ^2\left( \theta \right) \left( e^{-i\epsilon _2t}-e^{i\epsilon
_2t}\right)  & V\sin \left( \theta \right) \cos \left( \theta \right) \left(
e^{-i\epsilon _1t}-e^{i\epsilon _2t}\right)  \\ 
V\sin \left( \theta \right) \cos \left( \theta \right) \left( e^{-i\epsilon
_2t}-e^{i\epsilon _1t}\right)  & UV\sin ^2\left( \theta \right) \left(
e^{i\epsilon _1t}-e^{-i\epsilon _1t}\right) 
\end{array}
\right) .
\end{equation}
The condensates of the free-field particles are  
\begin{equation}
Z_1^{\prime }=Z_2^{\prime }=V^2\sin ^2\left( \theta \right) 
\end{equation}
and the condensates of the flavor particles are
\begin{equation}
\begin{array}{c}
Z_1=4V^2\sin ^2\left( \theta \right) \left( \cos ^2\left( \theta \right)
\sin ^2\left( 
\frac{\Omega _{12}t}2\right) +U^2\sin ^2\left( \theta \right) \sin ^2\left( 
\frac{\Omega _{22}t}2\right) \right) , \\ Z_2=4V^2\sin ^2\left( \theta
\right) \left( \cos ^2\left( \theta \right) \sin ^2\left( \frac{\Omega _{12}t
}2\right) +U^2\sin ^2\left( \theta \right) \sin ^2\left( \frac{\Omega _{11}t}
2\right) \right) .
\end{array}
\end{equation}
The vacuum structure is defined by the matrix $\hat Z$: 
$$
\hat Z=\frac {-1}{\cos ^2(\theta )+U^2\sin ^2(\theta )}\left( 
\begin{array}{cc}
-UV\sin ^2(\theta ) & V\cos (\theta )\sin (\theta ) \\ 
V\cos (\theta )\sin (\theta ) & UV\sin ^2(\theta )
\end{array}
\right)  
$$
with the normalization constant being ${\cal Z}=\frac 1{\cos ^2(\theta )+U^2\sin
^2(\theta )}=\frac 1{1-V^2\sin ^2(\theta )}$.

The time evolution of the flavor particle number (if \#1 was emitted) is
then given by: 
\begin{equation}
\begin{array}{c}
N_{111}=1-4U^2\sin ^2\left( \theta \right) \cos ^2\left( \theta \right) \sin
^2\left( 
\frac{\omega _{12}t}2\right) , \\ \bar N_{111}=4V^2\sin ^2\left( \theta
\right) \cos ^2\left( \theta \right) \sin ^2\left( \frac{\Omega _{12}t}2
\right)\ ,
\end{array}
\end{equation}
\begin{equation}
\begin{array}{c}
N_{121}=4\sin ^2\left( \theta \right) \{U^2\cos ^2\left( \theta \right) \sin
^2\left( 
\frac{\omega _{12}t}2\right) +V^2\cos ^2\left( \theta \right) \sin ^2\left( 
\frac{\Omega _{12}t}2\right) + \\ +U^2V^2\sin ^2\left( \theta \right) \sin
^2\left( 
\frac{\Omega _{11}t}2\right)\} , 
\\ \bar N_{121}=4U^2V^2\sin ^4\left( \theta
\right) \sin ^2\left( \frac{\Omega _{11}t}2\right) .
\end{array}
\end{equation}

\setcounter{equation}{0} \setcounter{figure}{0} 
\renewcommand{\theequation}{\mbox{5.\arabic{equation}}} 
\renewcommand{\thefigure}{\mbox{5.\arabic{figure}}}

\section{Conclusion}

The quantum field mixing effects may be understood by considering interplay
between the two Fock-spaces of the free-fields and the interacting fields. As
demonstrated in the 2-field mixing treatment, this interplay is highly
non-trivial and gives rise to a deviation from the simple quantum mechanical
approach due to the high-frequency oscillations and the
antiparticle componentû in the system.

We have now extended the previous results and presented a solution 
without approximations for the
quantum field theory of mixings in the arbitrary number of fields with
boson or fermion statistics. As one might have expected from the previous 
2-field treatment \cite{3,5,7,23}, all results fall into the same scheme 
and can be easily unified. We investigated the field time dynamics by 
calculating unequal time commutators and discussed the propagation 
functions. We found an explicit solution
for the interacting field Fock space and the corresponing vacuum structure
that turned out to be a generalized coherent state. We then showed the unitary
inequivalence between the mixed-field Fock space and the free-field Fock space in the
infinite volume limit. After we built a formal calculational framework, we
applied it to solve mixing dynamics of 2 vector mesons($S=1$) and 
fermions($S=1/2$). We found that
the scalar/pseudoscalar ($S=0$) boson mixing is the same as the mixing of 
transverse components of the 
vector fields, while for the longitudinal component of the vector field we 
found richer momentum dependence than in the spin-zero case.

However, from the application of our approach to 3-fermion/boson mixing cases,
which we summarize in Appendix C,
we saw very complicate structure of more general results. Oscillation formulas
typically involve all possible low-frequency and high-frequency combination
terms. The amplitudes of the oscillation terms are essentially momentum
dependent. We have also discussed the existence of the 
coherent antiparticle beam
generated from the starting definite-flavor particle beam and presented 
its dynamics.

Our general approach does not require to use any 
specific continuous parametrization of the mixing group but directly
takes the values of matrix elements. 
This allows an analysis to be carried out in a unified closed form
as shown in Sections II and III. In general, it may be preferable to solve 
the mixing problems without going through the intermediate parametrization 
step for the mixing matrix. Even if one wants to use a specific 
parametrization scheme for the mixing matrix, it is rather straightforward
to formulate our general framework into a symbolic calculation system, 
like maple or mathematica, and carry out extensive calculations involving 
mixing parameters in short period of time. Examples of such calculations 
are shown in Appendix C.

The physical application of the above formalism can be seen in investigating
the neutrino mixing, mixing of gauge vector bosons governed by the Weinberg
angle in the electroweak theory as well as vector mesons such as the $\rho $
and $\omega $. It seems also possible to apply these results to consider
nonperturbative quark-mixing effects in the Standard Model and provide
partial summation of the regular perturbation theory in mixing degrees of
freedom. For this purpose considering covariant form of the above theory
might be of great interest. Consideration along this line is in progress.

\acknowledgements
We thank Prof. G.Vitiello and Dr. M.Blasone for helpful discussions.
This work was supported by a grant from the U.S. Department of
Energy (DE-FG02-96ER 40947). 
The North Carolina Supercomputing Center and the National
Energy Research Scientific Computer Center are also acknowledged for the
grant of computing time.

\begin{appendix}
\setcounter{equation}{0} \setcounter{figure}{0} 
\renewcommand{\theequation}{\mbox{A.\arabic{equation}}} 
\renewcommand{\thefigure}{\mbox{A.\arabic{figure}}}

\section{Essential Cases of Mixing Field Parameters}

The most essential cases in modern particle physics are
scalar/pseudoscalar (spin 0),
vector (spin 1) boson fields and spin $\frac 12$ fermion fields. For these
cases mixing theory parameters are explicitly derived from quantum field
theory \cite{2,21}. We then have for scalar/pseudoscalar fields (spin 0):
\begin{equation}
u_{\vec k,0}=v_{\vec k,0}=1,
\end{equation}
for vector fields (spin 1):
\begin{equation}
\begin{array}{c}
u_{
\vec k,0}=v_{\vec k,0}=\left( \frac km,i\frac{\epsilon \left( k\right) }m
\vec n\right) , \\
u_{\vec k,\pm 1}=v_{\vec k,\pm 1}=\left( 0,i\vec n_{\pm}\right),
\end{array}
\end{equation}
where $\vec n=\frac{\vec k}k=\vec e_z$ and $\vec n_\pm=\mp\frac{1}{\sqrt{2}}
(\vec e_x\pm i \vec e_y )$ form a spherical basis.
For bispinor fields (spin $1/2$), we use the standard representation of the 
$\gamma$-matrices given by
\begin{equation}
\begin{array}{c}
\gamma^0=\left(
\begin{array}{cc}
\hat I & 0 \\
0 & -\hat I \\
\end {array} \right),
\vec \gamma=\left( \begin{array}{cc}
0 & \vec \sigma \\
-\vec \sigma & 0
\end{array} \right),
\end{array}
\end{equation}
and the corresponding representations of spinors:
\begin{equation}
\begin{array}{c}
u_{
\vec k,\sigma }=(\sqrt{\epsilon \left( k\right) +m}\omega _\sigma ,\sqrt{
\epsilon \left( k\right) -m}\left( \vec n\vec \sigma \right) \omega _\sigma
), \\ v_{-\vec k,\sigma }=(-\sqrt{\epsilon \left( k\right) -m}\left({\vec n}
\cdot{\vec \sigma} \right) \omega _{-\sigma },\sqrt{\epsilon \left( k\right)
+m} \omega _{-\sigma }),
\end{array}
\end{equation}
where $\omega _\sigma $ is spinor satisfying $\left( {\vec n}\cdot{\vec
\sigma}\right) \omega _\sigma =\sigma \cdot \omega _\sigma $ and $\sigma
$ takes values $\pm 1$.

The $H$ and $h$ matrix parameters are then for scalar case:
\begin{equation}
H^{\mu j}=h^{\mu j}=1,
\end{equation}
for spin 1:
\begin{equation}
\label{Hhboson}
\begin{array}{c}
\left\{
\begin{array}{c}
H_{
\vec k,0}^{\mu j}=\frac{\epsilon _\mu\left( k\right) \epsilon _j\left( 
k\right)
-k^2}{m_\mu m_j}, \\
h_{\vec k,0}^{\mu j}=\frac{\epsilon _\mu\left( k\right)
\epsilon _j\left( k\right) +k^2}{m_\mu m_j}
\end{array}
\right. ,\sigma =0 \\
H_{\vec k,\pm }^{\mu j}=h_{\vec k,\pm }^{\mu j}=1,\sigma =\pm 1;
\end{array}
\end{equation}
and for spin 1/2:
\begin{equation}
\label{Hhfermion}
\begin{array}{c}
H_{
\vec k,\sigma }^{\mu j}=\sqrt{\left( \epsilon _\mu\left( k\right) 
+m_\mu\right)
\left( \epsilon _j\left( k\right) +m_j\right) }+\sqrt{\left( \epsilon
_\mu \left( k\right) -m_\mu \right) \left( \epsilon _j\left( k\right) 
-m_j\right
) }, \\
h_{\vec k,\sigma }^{\mu j}=\sigma \left( \sqrt{\left( \epsilon _\mu\left(
k\right) -m_\mu\right) \left( \epsilon _j\left( k\right) +m_j\right) }-\sqrt{
\left( \epsilon _\mu\left( k\right) +m_\mu\right) \left( \epsilon _j\left(
k\right) -m_j\right) }\right) .
\end{array}
\end{equation}

\setcounter{equation}{0} \setcounter{figure}{0} 
\renewcommand{\theequation}{\mbox{B.\arabic{equation}}} 
\renewcommand{\thefigure}{\mbox{B.\arabic{figure}}}

\section{The flavor vacuum state}

In this appendix we explicitly solve flavor vacuum structure. We first
consider boson case. 

We write the sought flavor vacuum state as the most general linear combination
from the original-field Fock space 
\begin{equation}
|0>_f =\sum_{(n),(l)}\frac 1{n_1!n_2!\ldots n_k!}B_{\left(
n\right) \left( l\right) }\left( a_1^{\dagger }\right) ^{n_1}\ldots \left(
a_k^{\dagger }\right) ^{n_k}\left( b_{-1}^{\dagger }\right) ^{l_1}\ldots
\left( b_{-k}^{\dagger }\right) ^{l_k}|0>_m.
\end{equation}
From the particle/antiparticle symmetry, the part of Eq.(\ref{vacuumsolve})
involving antiparticle annihilation operators results in a dependent
set of equations and thus can be omitted. Expanding 
Eq.(\ref{vacuumsolve}), we find: 
\begin{equation}
\label{eqnset}\sum_j\left( \alpha _{ij}B_{\left( n_j+1\right) \left(
l\right) }+\beta _{ij}B_{\left( n\right) \left( l_j-1\right) }\right) =0,
\text{all }\left( n\right) ,\left( l\right) 
\end{equation}
where $(n_j+1)$ notation stands for $(n_1,n_2,\ldots ,n_j+1,\ldots n_k)$ and 
$k$ is the number of flavor fields. To solve this infinite set of equations
we introduce symbolic operators which decrease the subscript index of $B$
coefficients, i.e. $d_{-j}B_{\left( n\right) \left( l\right) }=B_{\left(
n\right) \left( l_j-1\right) }$. Then solving each set of equation in (\ref
{eqnset}) with respect to $B_{(n_j+1)(l)}$ we find 
\begin{equation}
\label{eqnstep}
\begin{array}{c}
B_{\left( n_i+1\right) \left( l\right)
}=(\sum\limits_jZ_{ij}d_{-j})B_{\left( n\right) \left( l\right) }
\text{ and consequently} 
\\ B_{\left( n\right) \left( l\right)}=
\prod\limits_i(\sum\limits_jZ_{ij}d_{-j})^{n_i}B_{\left( 0\right) \left(l 
\right) } \end{array}
\end{equation}
with matrix $\hat Z=-\hat \alpha ^{-1}\cdot \hat \beta $. 
Considering the momentum
conservation and the original Eq.(\ref{eqnset}), it
can be shown that only $B_{\left( 0\right) \left( l=0\right) }$ must be
non-zero among all $\left( l\right) $. Thus,
applying symbolic operators $d_{-j}$ and leaving only terms 
$B_{\left( 0\right) \left( 0\right) }$ in the expansion, we get 
\begin{equation}
\label{vac01}
B_{\left( n\right) \left( l\right) }=\sum_{
\left\{ 
\begin{array}{c}
(j_p^i) \\ 
\sum\limits_pj_p^i=n_i \\ 
\sum\limits_ij_p^i=l_p
\end{array}
\right\} }\prod_i\frac{n_i!}{j_1^i!\ldots j_k^i!}Z_{i1}^{j_1^i}\ldots
Z_{ik}^{j_k^i}B_{\left( 0\right) \left( 0\right) }.
\end{equation}
It is possible to rewrite this complicate expression in the more compact 
form; 
\begin{equation}
\label{compvac}|0>_f =\frac 1{{\cal Z}}\sum_{\left( k\right) }\prod_i\frac 
1{k_i!}(\sum_jZ_{ij}a_i^{\dagger }b_{-j}^{\dagger })^{k_i}|0>_m
\end{equation}
that can be shown directly by expanding the above expression. It also can be
argued that to obtain $B_{\left( n\right) \left( l\right) }$ from Eq.(\ref
{compvac}) one needs to leave only those terms in the expansion that give 
correct
power of particle and antiparticle creation operators, i.e. total powers of all 
$a_i^{\dagger }$'s are $n_i$'s and $b_i^{\dagger}$'s are $l_i$. 
But this is same with extracting $B_{\left(n\right) \left( l\right) }$ 
from Eq.(\ref{eqnstep}). The constant ${\cal Z}$
is introduced instead of $B_{\left( 0\right) \left( 0\right) }$ and serves
as a normalization factor determined by $_f<0|0>_f =1$.

The Eq.(\ref{compvac}) can be further simplified as 
\begin{equation}
\begin{array}{rl}
|0>_f &=
\frac 1{\cal Z}\sum\limits_{\left( k\right) }\prod\limits_i\frac 1{k_i!}
(\sum\limits_jZ_{ij}a_i^{\dagger }b_{-j}^{\dagger })^{k_i}|0>_m= 
\\ &=\frac 1{\cal Z}\prod\limits_i\sum\limits_{k_i=0}^\infty \frac 1{k_i!}
(\sum\limits_jZ_{ij}a_i^{\dagger }b_{-j}^{\dagger })^{k_i}|0>_m=
\\ &=\frac 1{\cal Z}\exp (\sum\limits_{i,j=1}^NZ_{ij}a_i^{\dagger
}b_{-j}^{\dagger })|0>_m .
\end{array}
\end{equation}

Let us now proceed to the fermion case. We employ the same idea 
with the symbolic shifting operators. If $\hat C
_{\left( n\right) \left( l\right) }$ stands for creation operator for
fermion state $|(n),(l)>$, we want then 
\begin{equation}
\begin{array}{c}
a_iB_{(n_i+1)(l)}
\hat C_{\left( n_i+1\right) \left( l\right) }|0>_m=\pm B_{(n_i+1)(l)}\hat C
_{\left( n\right) \left( l\right) }|0>_m=d_{+i}B_{(n)(l)}\hat C_{\left(
n\right) \left( l\right) }|0>_m \\ b_i^{\dagger }B_{(n)(l_i-1)}\hat C
_{\left( n\right) \left( l_i-1\right) }|0>_m=\pm B_{(n)(l_i-1)}\hat C
_{\left( n\right) \left( l\right) }|0>_m=d_{-i}B_{(n)(l)}\hat C_{\left(
n\right) \left( l\right) }|0>_m
\end{array}
\end{equation}
with correct sign. Eq.(\ref{eqnset}) then can be written in the form 
\begin{equation}
\sum_j(\alpha _{ij}d_{+j}+\beta _{ij}d_{-j})B_{\left( n\right) \left(
l\right) }=0
\end{equation}
which binds together the shifting operators that increase and decrease 
the index. This set can be solved as 
\begin{equation}
d_{+i}[B_{\left( n\right) \left( l\right) }]=\sum_jZ_{ij}d_{-j}[B_{\left(
n\right) \left( l\right) }]
\end{equation}
with the same matrix $\hat{Z}$ presented in the boson case. From the 
definition of 
shifting operators it can be inferred that they obey anticommutation 
property (i.e. 
$d_{\pm i}d_{\pm j}=-d_{\pm j}d_{\pm i}$) and thus it can be shown further 
that for $i_1>i_2>\ldots >i_n$
\begin{equation}
\begin{array}{c}
d_{+i_n}d_{+i_{n-1}}\ldots d_{+i_1}B_{\left( 0\right) \left( l\right)
}=B_{\left( i\right) \left( l\right) } \\ 
d_{-i_1}d_{-i_2}\ldots d_{-i_l}B_{\left( n\right) \left( l\right)
}=B_{\left( n\right) \left( l-i\right) }
\end{array}
\end{equation}
so that the solution can be written again as
\begin{equation}
B_{\left( n\right) \left( l\right)
}=\prod_i(\sum_jZ_{ij}d_{-j})^{n_i}B_{\left( 0\right) \left( l\right) },
\end{equation}
where only $B_{\left( 0\right) \left( 0\right) }$ survives. Here, $n_i$ 
can be only 0 or 1 and the anticommutation rules for the ordering are 
applied. It is remarkable that
Eq.(\ref{compvac}) can still be used for the fermion vacuum. This can be
verified by a direct expansion with the anticommutation nature of
ladder operators. Thus, for either boson or fermion case the flavor 
vacuum state can be written as 
\begin{equation}
|0>_f =\frac 1{{\cal Z}}\exp (\sum_{i,j=1}^NZ_{ij}a_i^{\dagger
}b_{-j}^{\dagger })|0>_m.
\end{equation}
We now proceed to find the normalization constant ${\cal Z}$. To do this, we 
consider \begin{equation}
||0>_f |^2=|\exp (\sum_{i,j=1}^NZ_{ij}a_i^{\dagger }b_{-j}^{\dagger
})|0>_m|^2=\sum_L\frac 1{L!^2}|(\sum_{i,j=1}^NZ_{ij}a_i^{\dagger
}b_{-j}^{\dagger })^L|0>_m|^2,
\end{equation}
where we use the fact that the states of 
$(\sum_{i,j=1}^NZ_{ij}a_i^{\dagger }b_{-j}^{\dagger})^L|0>_m$ 
are orthogonal for different $L$'s. We then employ the
fact that matrix $\hat Z$ can be transformed to a diagonal form with two
unitary transformations, {\it i.e.}
\begin{equation}
Z^{\prime }=\left( 
\begin{array}{ccc}
x_1 & 0 & \ldots  \\ 
0 & \ddots  & 0 \\ 
\ldots  & 0 & x_N
\end{array}
\right) =UZV^{\dagger }.
\end{equation}
We can now introduce additional unitary transformations of $\ a^{\prime
\dagger }=U^{\dagger }a^{\dagger }$, $b^{\prime \dagger }=V^{\dagger
}b^{\dagger }$ to make $\sum\limits_{i,j=1}^NZ_{ij}a_i^{\dagger
}b_{-j}^{\dagger }=\sum\limits_{i=1}^NZ_{ii}^{\prime }a_i^{\prime \dagger
}b_{-i}^{\prime \dagger }$, where $a_i^{\prime },b_{-j}^{\prime }$ satisfy
the standard commutation/anticommutation relationship. Then, using 
the binomial formula to expand $(\sum\limits_{i=1}^NZ_{ii}^{\prime 
}a_i^{\prime \dagger }b_{-i}^{\prime \dagger })^L$, we find 
\begin{equation}
\label{normsum}
\begin{array}{c}
\sum\limits_L
\frac 1{L!^2}|(\sum\limits_{i=1}^NZ_{ii}^{\prime }a_i^{\prime \dagger
}b_{-i}^{\prime \dagger })^L|0>_m|^2= \\ =\sum\limits_L
\frac 1{L!^2}\sum\limits_{n_1+\ldots +n_N=L}L!^2\prod\limits_{j=1}^N\frac 1{
n_j!^2}|(Z_{jj}^{\prime }a_j^{\prime \dagger }b_{-j}^{\prime \dagger
})^{n_j}|0>_m|^2= \\ 
=\sum\limits_L\sum\limits_{n_1+\ldots
+n_N=L}\prod\limits_{j=1}^N\frac{n_j!^2}{n_j!^2}|Z_{jj}^{\prime n_j}|^2=
\sum\limits_{n_1,\ldots ,n_N}\lambda _1^{n_1}\ldots \lambda _N^{n_N},
\end{array}
\end{equation}
where $\lambda _i$'s are eigenvalues of $ZZ^{\dagger }$. The summation
limits in Eq.(\ref{normsum}) are different for fermions and bosons. For
bosons $n_i$ runs from $0$ to $\infty $, while for fermions they only
can be $0$ or $1$. In either case the sum can be evaluated to give 
\begin{equation}
|\exp (\sum_{i,j=1}^NZ_{ij}a_i^{\dagger }b_{-j}^{\dagger })|0>_m|^2=\left\{ 
\begin{array}{c}
\prod\limits_i\left( 1+\lambda _i\right) 
\text{ fermions} \\ \prod\limits_i\frac 1{1-\lambda _i}\text{ bosons }
\end{array}
\right. =\left\{ 
\begin{array}{c}
\det (
\hat 1+ZZ^{\dagger })\text{ fermions} \\ \det \nolimits^{-1}(\hat 1
-ZZ^{\dagger })\text{ bosons }
\end{array}
\right. .
\end{equation}

\setcounter{equation}{0} \setcounter{figure}{0} 
\renewcommand{\theequation}{\mbox{C.\arabic{equation}}} 
\renewcommand{\thefigure}{\mbox{C.\arabic{figure}}}

\section{Unitary mixing of 3 fields in Wolfenstein parametrization}

We now present application of the above general formalism to specific case
of mixing of 3 quantum fields. Calculations were carried out with the help of
Mathematica 3 symbolic calculational system.

We note that all time-dependent quantities in this section are presented in
the form of matrices each entry of which corresponds to certain $\Omega
_{ij}=\omega _i+\omega _j$ or $\omega _{ij}=\omega _i-\omega _j$ frequency. 
It means that each quantity is presented in the form 
\begin{equation}
P=2Re\left( \sum_{ij}\left[ P_{ij}^\Omega e^{-i\Omega _{ij}t}+P_{ij}^\omega
e^{-i\omega _{ij}t}\right] \right),
\end{equation}
where 
$P^\Omega $ and $P^\omega $
matrices are written as follows: 
\begin{eqnarray}
P^\Omega =\{\{P_{11}^\Omega ,P_{12}^\Omega ,P_{13}^\Omega \},\{P_{21}^\Omega
,P_{22}^\Omega ,P_{23}^\Omega \},\{P_{31}^\Omega ,P_{32}^\Omega
,P_{33}^\Omega \}\}\cr
P^\omega =\{\{P_{11}^\omega ,P_{12}^\omega ,P_{13}^\omega \},\{P_{21}^\omega
,P_{22}^\omega ,P_{23}^\omega \},\{P_{31}^\omega ,P_{32}^\omega
,P_{33}^\omega \}\}.
\end{eqnarray}
Since the diagonal elements of $P^\omega$ corresponds to the same 
constant term $\omega _{ii}=0$, 
we can collect the diagonal elements of $P^\omega$ as
$Sp\left(P^\omega \right) = P_{11}^\omega + P_{22}^\omega + P_{33}^\omega$ 
and express only the off-diagonal elements as 
\begin{eqnarray}
{\tilde P}^\omega =\{\{0 ,P_{12}^\omega ,P_{13}^\omega 
\},\{P_{21}^\omega, 0 ,P_{23}^\omega \},\{P_{31}^\omega ,P_{32}^\omega
, 0 \}\}.
\end{eqnarray}

\subsection{The case of 3 fermion fields}

We now show our results for the unitary mixing of 3 fields with spin 
$\frac 12$(bispinors). Although an explicit parametrization is not needed 
in our formalism, we may use Wolfenstein parametrization as an explicit form 
of a mixing matrix 
\begin{equation}
U=\left( 
\begin{array}{ccc}
1-\lambda ^2/2 & \lambda  & A\lambda ^3\left( \rho -i\eta \right)  \\ 
-\lambda  & 1-\lambda ^2/2 & A\lambda ^2 \\ 
A\lambda ^3\left( 1-\rho -i\eta \right)  & -A\lambda ^2 & 1
\end{array}
\right) .
\end{equation}
All results are then computed to a few lowest orders in $\lambda$. 

For the bispinors, we redefine our $H$ and $h$ matrices as 
$H^{ij}\rightarrow H^{ij}/\left( 2\sqrt{\epsilon _i\epsilon _j}\right) $, 
$h^{ij}\rightarrow h^{ij}/\left( 2\sqrt{\epsilon _i\epsilon _j}\right) $ 
so that 
\begin{equation}
\begin{array}{c}
H=\left( 
\begin{array}{ccc}
1 & u_{12} & u_{13} \\ 
u_{12} & 1 & u_{23} \\ 
u_{13} & u_{23} & 1
\end{array}
\right) ; \\ 
h=\left( 
\begin{array}{ccc}
0 & v_{12} & v_{13} \\ 
-v_{12} & 0 & v_{23} \\ 
-v_{13} & -v_{23} & 0
\end{array}
\right) .
\end{array}
\end{equation}
Also, $u_{ij},v_{ij}$ are defined in the same way as in the 2 field mixing
\begin{equation}
\begin{array}{c}
u_{ij}=
\frac{\sqrt{\left( \epsilon _i+m_i\right) \left( \epsilon _j+m_j\right) }+
\sqrt{\left( \epsilon _i-m_i\right) \left( \epsilon _j-m_j\right) }}{2\sqrt{
\epsilon _i\epsilon _j}}, 
\\ v_{ij}=\sigma \frac{\sqrt{\left( \epsilon
_i-m_i\right) \left( \epsilon _j+m_j\right) }-\sqrt{\left( \epsilon
_i+m_i\right) \left( \epsilon _j-m_j\right) }}{2\sqrt{\epsilon _i\epsilon _j}
}.
\end{array}
\end{equation}
Then, the structure of the ladder operators is described by $\alpha $ and 
$\beta $ matrices 
\begin{equation}
\begin{array}{c}
\alpha =\left( 
\begin{array}{ccc}
1-\lambda ^2/2 & u_{12}\lambda  & u_{13}A\lambda ^3\left( \rho -i\eta
\right)  \\ 
-u_{12}\lambda  & 1-\lambda ^2/2 & u_{23}A\lambda ^2 \\ 
u_{13}A\lambda ^3\left( 1-\rho -i\eta \right)  & -u_{23}A\lambda ^2 & 1
\end{array}
\right) , \\ 
\beta =\left( 
\begin{array}{ccc}
0 & v_{12}\lambda  & v_{13}A\lambda ^3\left( \rho -i\eta \right)  \\ 
v_{12}\lambda  & 0 & v_{23}A\lambda ^2 \\ 
-v_{13}A\lambda ^3\left( 1-\rho -i\eta \right)  & v_{23}A\lambda ^2 & 0
\end{array}
\right) .
\end{array}
\end{equation}
To make the results more compact, we define $c=A\left( \rho -i\eta \right) 
$, $ e=-A\left( 1-\rho -i\eta \right) $ and $a=A$ so that 
\begin{equation}
\label{eq01}
\begin{array}{c}
\alpha =\left( 
\begin{array}{ccc}
1-\lambda ^2/2 & u_{12}\lambda  & u_{13}c\lambda ^3 \\ 
-u_{12}\lambda  & 1-\lambda ^2/2 & u_{23}a\lambda ^2 \\ 
-u_{13}e\lambda ^3 & -u_{23}a\lambda ^2 & 1
\end{array}
\right) , \\ 
\beta =\left( 
\begin{array}{ccc}
0 & v_{12}\lambda  & v_{13}c\lambda ^3 \\ 
v_{12}\lambda  & 0 & v_{23}a\lambda ^2 \\ 
v_{13}e\lambda ^3 & v_{23}a\lambda ^2 & 0
\end{array}
\right) .
\end{array}
\end{equation}

For the case when \#2 flavor particle
was initially present, the flavor charge oscillation formulas are as 
follows. The flavor charge fluctuation, $Q_{212}(t)$, is given by 
\begin{equation}
\begin{array}{c}
Q_{212}^\Omega =\{\{0,-
\frac{\lambda ^2v_{12}^2}2\left( 1-\lambda ^2\right) ,-\frac{\lambda
^6av_{13}c^{*}}2\left( u_{23}v_{12}+u_{12}v_{23}\right) \}, \\ \{-
\frac{\lambda ^2v_{12}^2}2\left( 1-\lambda ^2\right) ,0,\frac{\lambda
^6av_{23}}2\left( u_{12}v_{13}c^{*}-u_{13}v_{12}c\right) \}, \\ \{-\frac{
\lambda ^6av_{13}c^{*}}2\left( u_{23}v_{12}+u_{12}v_{23}\right) ,\frac{
\lambda ^6av_{23}}2\left( u_{12}v_{13}c^{*}-u_{13}v_{12}c\right) ,0\}\},
\end{array}
\end{equation}
\begin{equation}
\begin{array}{c}
{\tilde Q}_{212}^\omega =\{\{0,-
\frac{\lambda ^2u_{12}^2}2\left( 1-\lambda ^2\right) ,\frac{\lambda
^6au_{13}c}2\left( -u_{12}u_{23}+v_{12}v_{23}\right) \}, \\ \{-
\frac{\lambda ^2u_{12}^2}2\left( 1-\lambda ^2\right) ,0,\frac{\lambda
^6au_{23}}2\left( u_{12}u_{13}c+v_{12}v_{13}c^{*}\right) \}, \\ \{
\frac{\lambda ^6au_{13}c^{*}}2\left( -u_{12}u_{23}+v_{12}v_{23}\right) ,
\frac{\lambda ^6au_{23}}2\left( u_{12}u_{13}c^{*}+v_{12}v_{13}c\right)
,0\}\}, \\ Sp\left( Q_{212}^\omega \right) =\lambda ^2\left(
u_{12}^2+v_{12}^2\right) \left( 1-\lambda ^2\right) .
\end{array}
\end{equation}
Similarly, $Q_{222}(t)$ and $Q_{232}(t)$ are given by 
\begin{equation}
\begin{array}{c}
Q_{222}^\Omega =\{\{0,
\frac{v_{12}^2\lambda ^2}2\left( 1-\lambda ^2\right) ,\frac{a^2\left(
u_{23}v_{12}+u_{12}v_{23}\right) ^2}2\lambda ^6\}, \\ \{
\frac{v_{12}^2\lambda ^2}2\left( 1-\lambda ^2\right) ,0,\frac{
a^2v_{23}^2\lambda ^4}2\left( 1-\lambda ^2\right) \}, \\ \{\frac{a^2\left(
u_{23}v_{12}+u_{12}v_{23}\right) ^2}2\lambda ^6,\frac{a^2v_{23}^2\lambda ^4}2
\left( 1-\lambda ^2\right) ,0\}\},
\end{array}
\end{equation}
\begin{equation}
\begin{array}{c}
{\tilde Q}_{222}^\omega =\{\{0,
\frac{u_{12}^2\lambda ^2}2\left( 1-\lambda ^2\right) ,\frac{a^2\left(
u_{23}u_{12}-v_{12}v_{23}\right) ^2}2\lambda ^6\}, \\ \{
\frac{u_{12}^2\lambda ^2}2\left( 1-\lambda ^2\right) ,0,\frac{
a^2u_{23}^2\lambda ^4}2\left( 1-\lambda ^2\right) \}, \\ \{
\frac{a^2\left( u_{23}u_{12}-v_{12}v_{23}\right) ^2}2\lambda ^6,\frac{
a^2u_{23}^2\lambda ^4}2\left( 1-\lambda ^2\right) ,0\}\}, \\ Sp\left(
Q_{222}^\omega \right) =\frac 12-\lambda ^2+\frac 14\left(
3+2u_{12}^4+4u_{12}^2v_{12}^2+2v_{12}^4\right) \lambda ^4,
\end{array}
\end{equation}

and 

\begin{equation}
\begin{array}{c}
Q_{232}^\Omega =\{\{0,
\frac{av_{12}\lambda ^6}2\left( eu_{13}v_{23}-u_{23}v_{13}e^{*}\right) ,
\frac{\lambda ^6av_{13}e^{*}}2\left( u_{23}v_{12}+u_{12}v_{23}\right) \}, \\ 
\{
\frac{av_{12}\lambda ^6}2\left( eu_{13}v_{23}-u_{23}v_{13}e^{*}\right) ,0,-
\frac{a^2v_{23}^2\lambda ^4}2+\frac{a^2v_{23}^2\lambda ^6}4\}, \\ \{\frac{
\lambda ^6av_{13}e^{*}}2\left( u_{23}v_{12}+u_{12}v_{23}\right) ,-\frac{
a^2v_{23}^2\lambda ^4}2+\frac{a^2v_{23}^2\lambda ^6}4,0\}\},
\end{array}
\end{equation}
\begin{equation}
\begin{array}{c}
{\tilde Q}_{232}^\omega =\{\{0,-
\frac{au_{12}\lambda ^6}2\left( ev_{13}v_{23}+u_{13}u_{23}e^{*}\right) ,
\frac{\lambda ^6au_{13}e^{*}}2\left( u_{12}u_{23}-v_{12}v_{23}\right) \}, \\ 
\{-
\frac{au_{12}\lambda ^6}2\left( e^{*}v_{13}v_{23}+u_{13}u_{23}e\right) ,0,-
\frac{a^2u_{23}^2\lambda ^4}2+\frac{a^2u_{23}^2\lambda ^6}4\}, \\ \{
\frac{\lambda ^6au_{13}e}2\left( u_{12}u_{23}-v_{12}v_{23}\right) ,-\frac{
a^2u_{23}^2\lambda ^4}2+\frac{a^2u_{23}^2\lambda ^6}4,0\}\}, \\ Sp\left(
Q_{232}^\omega \right) =a^2\left( u_{23}^2+v_{23}^2\right) \lambda ^4\left(
1-\lambda ^2/2\right),
\end{array}
\end{equation}
respectively.

In more details, the dynamics is given by the following quantities. 
The non-equal time anticommutators are given by: 
\begin{equation}
\begin{array}{c}
F_{11}(t)=e^{-i\epsilon _1t}+\lambda ^2\left( -e^{-i\epsilon
_1t}+u_{12}^2e^{-i\epsilon _2t}+v_{12}^2e^{i\epsilon _2t}\right) , \\ 
F_{12}(t)=F_{21}(t)=\lambda u_{12}\left( e^{-i\epsilon
_2t}-e^{-i\epsilon _1t}\right) +\lambda ^3
\frac{u_{12}}2\left( e^{-i\epsilon _1t}-e^{-i\epsilon _2t}\right) , \\ 
F_{13}(t)=F_{31}(-t)^{*}=
\lambda^3 \left(
u_{13}(c e^{-i\epsilon _3t}-e^{*}e^{-i\epsilon _1t})
-au_{12}u_{23}e^{-i\epsilon_2t}+av_{12}v_{23}e^{i\epsilon _2t}-\right)  \\ 
F_{22}(t)=e^{-i\epsilon _2t}+\lambda ^2\left( -e^{-i\epsilon
_2t}+u_{12}^2e^{-i\epsilon _1t}+v_{12}^2e^{i\epsilon _1t}\right) , \\ 
F_{23}(t)=F_{32}(t)=\lambda ^2au_{23}\left(
e^{-i\epsilon _3t}-e^{-i\epsilon _2t}\right) , \\ 
F_{33}=e^{-i\epsilon _3t};
\end{array}
\end{equation}
\begin{equation}
\begin{array}{c}
G_{11}(t)=\lambda ^2u_{12}v_{12}\left( e^{-i\epsilon _2t}-e^{i\epsilon
_2t}\right) , \\ 
G_{12}(t)=-(G_{21}(t))^{*}=\lambda v_{12}\left( e^{-i\epsilon
_1t}-e^{i\epsilon _2t}\right) +\lambda ^3
\frac{v_{12}}2\left( -e^{-i\epsilon _1t}+e^{i\epsilon _2t}\right) , \\ 
G_{13}(t)=-(G_{31}(t))^{*}=\lambda ^3\left( 
v_{13}(e e^{-i\epsilon _1t}-c^{*}e^{i\epsilon _3t})+
au_{23}v_{12}e^{i\epsilon_2t}+au_{12}v_{23}e^{-i\epsilon_2t}\right) , \\ 
G_{22}(t)=\lambda ^2u_{12}v_{12}\left( -e^{-i\epsilon _1t}+e^{i\epsilon
_1t}\right) , \\ 
G_{23}(t)=-(G_{32}(t))^{*}=\lambda ^2av_{23}\left( e^{-i\epsilon
_2t}-e^{i\epsilon _3t}\right) , \\ 
G_{33}(t)=\lambda ^4a^2u_{23}v_{23}\left( e^{i\epsilon _2t}-e^{-i\epsilon
_2t}\right) .
\end{array}
\end{equation}
The vacuum structure is defined by $\hat Z$ matrix: 
\begin{equation}
\label{eq02}
\begin{array}{c}
Z_{11}=u_{12}v_{12}\lambda ^2+u_{12}\left( 1-u_{12}^2\right) v_{12}\lambda
^4, \\ 
Z_{12}=-v_{12}\lambda -\left( 
\frac 12-u_{12}^2\right) v_{12}\lambda ^3=Z_{21}, \\ 
Z_{13}=-\left(
cv_{13}-au_{12}v_{23}\right) \lambda ^3,Z_{31}=-\left(
au_{23}v_{12}+ev_{13}\right) \lambda ^3, \\ 
Z_{22}=-Z_{11}+a^2u_{23}v_{23}\lambda ^4, \\ 
Z_{23}=-av_{23}\lambda ^2-\left( cu_{12}v_{13}+a\left( 
\frac 12-u_{12}^2\right) v_{23}\right) \lambda ^4, \\ 
Z_{32}=-av_{23}\lambda
^2-\left( eu_{13}+au_{12}u_{23}\right) v_{12}\lambda ^4, \\ 
Z_{33}=-a^2u_{23}v_{23}\lambda ^4.
\end{array}
\end{equation}
The normalization constant is obtained as ${\cal Z}\approx 1+v_{12}^2\lambda
^2+(v_{12}^2+a^2v_{23}^2-v_{12}^2u_{12}^2)\lambda ^4+\ldots $.

If the particle of sort \#2 was originally present, then for the 
particle of sort \#1 the mixing quantities are as follows. 
The free-field particle condensate is 
\begin{equation}
Z_1^{\prime }=v_{12}^2\lambda ^2
\end{equation}
and the flavor particle condensate is
\begin{equation}
\label{eq03}
\begin{array}{c}
Z_1^\Omega =\{\{0,-\lambda ^2
\frac{v_{12}^2}2(1-\lambda ^2),-\frac{\lambda ^6}2v_{13}c^{*}\left(
au_{23}v_{12}+v_{13}e^{*}\right) \}, \\ 
\{-\lambda ^2\frac{v_{12}^2}2(1-\lambda ^2),-\lambda ^4u_{12}^2v_{12}^2,
-\frac{\lambda ^6av_{23}}2\left(cu_{13}v_{12}+u_{12}v_{13}c^{*}\right) \}, \\ 
\{-\frac{\lambda ^6}2v_{13}c^{*}\left( au_{23}v_{12}+v_{13}e^{*}\right) 
,-\frac{
\lambda ^6}2av_{23}\left( cu_{13}v_{12}+u_{12}v_{13}c^{*}\right) , \\ 
-a^2cu_{13}u_{23}v_{13}v_{23}c^{*}\lambda ^{10}\}\},
\end{array}
\end{equation}
\begin{equation}
\label{eq04}
\begin{array}{c}
{\tilde Z}_1^\omega =\{\{0,
\frac{aeu_{12}v_{13}v_{23}}2\lambda ^6,\frac{acu_{13}v_{12}v_{23}}2\lambda
^6\},\{\frac{ae^{*}u_{12}v_{13}v_{23}}2\lambda ^6,0, \\ \frac{\lambda
^8v_{12}v_{13}c^{*}}2\left( 2cu_{12}u_{13}-\frac 12au_{23}\right) \},\{\frac{
ac^{*}u_{13}v_{12}v_{23}}2\lambda ^6, \\ \frac{\lambda ^8v_{12}v_{13}c}2
\left( 2c^{*}u_{12}u_{13}-\frac 12au_{23}\right) ,0\}\}; \\ Sp\left(
Z_1^\omega \right) =\lambda ^2v_{12}^2-\left( 1-u_{12}^2\right)
v_{12}^2\lambda ^4.
\end{array}
\end{equation}
The flavor particle number fluctuations are given by $N_{212}(t)=\left|
F_{12}(t)\right| ^2+Z_1(t)$: 
\begin{equation}
\begin{array}{c}
N_{212}^\Omega =\{\{-\lambda ^4u_{12}^2v_{12}^2,-\lambda ^2
\frac{v_{12}^2}2\left( 1-\lambda ^2\right) , \\ 
\frac{\lambda ^6a}2\left(
v_{13}v_{23}u_{12}\left( e^{*}-c^{*}\right)
-v_{12}v_{13}u_{23}c^{*}+2av_{12}v_{23}u_{12}u_{23}\right) \}, \\ 
\{-\lambda^2
\frac{v_{12}^2}2\left( 1-\lambda ^2\right) ,0,-\frac{a^2v_{23}^2}2\lambda ^4 \}, \\ 
\{\frac{\lambda ^6a}2\left( v_{13}v_{23}u_{12}\left( e^{*}-c^{*}\right)
-v_{12}v_{13}u_{23}c^{*}+2av_{12}v_{23}u_{12}u_{23}\right) , \\ 
-\frac{a^2v_{23}^2}2\lambda ^4 ,-a^4u_{23}^2v_{23}^2\lambda ^8\}\};
\end{array}
\end{equation}
\begin{equation}
\begin{array}{c}
{\tilde N}_{212}^\omega =\{\{0,-\lambda ^2u_{12}^2\left( 1-\lambda ^2\right) 
, \\ \frac{\lambda ^6a}2\left( v_{12}v_{23}\left(
-u_{13}e^{*}+u_{13}c-2au_{12}u_{23}\right) -cu_{12}u_{13}u_{23}\right)
\},\{-\lambda ^2u_{12}^2\left( 1-\lambda ^2\right) , \\ 
0,\frac{\lambda ^6au_{23}}2\left( cu_{12}u_{13}+c^{*}v_{12}v_{13}\right) \},
\\ \{
\frac{\lambda ^6a}2\left( v_{12}v_{23}\left(
-u_{13}e+u_{13}c^{*}-2au_{12}u_{23}\right) -c^{*}u_{12}u_{13}u_{23}\right) ,
\\ \frac{\lambda ^6au_{23}}2\left( c^{*}u_{12}u_{13}+cv_{12}v_{13}\right)
,0\}\}; \\ Sp\left( N_{212}^\omega \right) =\lambda ^2\left(
u_{12}^2+v_{12}^2\right) +\lambda ^4\left(
a^2v_{23}^2-v_{12}^2-u_{12}^2\left( 1-v_{12}^2\right) \right) ,
\end{array}
\end{equation}
and the flavor antiparticle number fluctuations, $\bar N_{212}(t)$,
are given by
\begin{equation}
\label{eq05}
\begin{array}{c}
\bar N_{212}^\Omega =\{\{-u_{12}^2v_{12}^2\lambda ^4,\frac{\lambda
^6aeu_{13}v_{12}v_{23}}2,\frac{\lambda ^6au_{12}v_{23}}2\left(
2au_{23}v_{12}+v_{13}e^{*}\right) \}, \\ 
\{\frac{\lambda ^6aeu_{13}v_{12}v_{23}}2,0,-\frac{a^2v_{23}^2\lambda ^4}2\}, \\ 
\{\frac{\lambda ^6au_{12}v_{23}}2\left(
2au_{23}v_{12}+v_{13}e^{*}\right) ,-\frac{a^2v_{23}^2\lambda ^4}2,
-a^4u_{23}^2v_{23}^2\lambda ^8\}\};
\end{array}
\end{equation}
\begin{equation}
\label{eq06}
\begin{array}{c}
{\tilde {\bar N}}_{212}^\omega =\{\{0,-\frac{aeu_{12}v_{13}v_{23}\lambda 
^6}2,-\frac{
av_{12}v_{23}\lambda ^6}2\left( 2au_{12}u_{23}+u_{13}e^{*}\right) \}, \\ \{-
\frac{ae^{*}u_{12}v_{13}v_{23}\lambda ^6}2,0,0\},\{-\frac{
av_{12}v_{23}\lambda ^6}2\left( 2au_{12}u_{23}+u_{13}e\right) ,0,0\}\}; \\ 
Sp\left( \bar N_{212}^\omega \right) =\lambda ^4\left(
u_{12}^2v_{12}^2+a^2v_{23}^2\right) .
\end{array}
\end{equation}

In the same initial condition, we obtain 
the following for the particle of sort \#2. The free-field particle 
condensate is  
\begin{equation}
Z_2^{\prime }=v_{12}^2\lambda ^2+a^2v_{23}^2\lambda ^4
\end{equation}
and the flavor particle condensate is
\begin{equation}
\label{eq07}
\begin{array}{c}
Z_2^\Omega =\{\{-u_{12}^2v_{12}^2\lambda ^4,-
\frac{v_{12}^2\lambda ^2}2\left( 1-\lambda ^2\right) , \\ \frac{\lambda ^6a}2
\left(
2au_{12}u_{23}v_{12}v_{23}-c^{*}u_{23}v_{12}v_{13}+e^{*}u_{12}v_{13}v_{23}
\right) \}, \\ \{-
\frac{v_{12}^2\lambda ^2}2\left( 1-\lambda ^2\right) ,0,-\frac{a^2v_{23}^2}2
\lambda ^4\}, \\ \{
\frac{\lambda ^6a}2\left(
2au_{12}u_{23}v_{12}v_{23}-c^{*}u_{23}v_{12}v_{13}+e^{*}u_{12}v_{13}v_{23}
\right) , \\ -\frac{a^2v_{23}^2}2\lambda ^4,-a^4u_{23}^2v_{23}^2\lambda
^8\}\};
\end{array}
\end{equation}
\begin{equation}
\label{eq08}
\begin{array}{c}
{\tilde Z}_2^\omega =\{\{0,-
\frac{aeu_{12}v_{13}v_{23}\lambda ^6}2,\frac{av_{12}v_{23}\lambda ^6}2\left(
cu_{13}-2au_{12}u_{23}-e^{*}u_{13}\right) \}, \\ \{-
\frac{ae^{*}u_{12}v_{13}v_{23}\lambda ^6}2,0,\frac{au_{23}v_{12}v_{13}c^{*}
\lambda ^6}2\}, \\ \{
\frac{av_{12}v_{23}\lambda ^6}2\left(
c^{*}u_{13}-2au_{12}u_{23}-eu_{13}\right) ,\frac{au_{23}v_{12}v_{13}c\lambda
^6}2,0\}\}; \\ Sp\left( Z_2^\omega \right) =v_{12}^2\lambda ^2+\left(
a^2v_{23}^2+v_{12}^2\left( u_{12}^2-1\right) \right) \lambda ^4.
\end{array}
\end{equation}
The flavor particle number fluctuations, $N_{222}(t)=\left|
F_{22}(t)\right| ^2+Z_2(t)$, are given by 
\begin{equation}
\begin{array}{c}
N_{222}^\Omega =\{\{-eu_{12}u_{13}v_{12}v_{13}e^{*}\lambda ^8,
\frac{aeu_{13}v_{12}v_{23}}2\lambda ^6, \\ \frac{\lambda ^6a}2\left( a\left(
u_{23}v_{12}+u_{12}v_{23}\right)
^2-u_{23}v_{12}v_{13}c^{*}+u_{12}v_{13}v_{23}e^{*}\right) \}, \\ \{
\frac{aeu_{13}v_{12}v_{23}}2\lambda ^6,0,-\frac{\lambda ^6av_{23}}2\left(
cu_{13}v_{12}+\frac{av_{23}}2\right) \}, \\ \{
\frac{\lambda ^6a}2\left( a\left( u_{23}v_{12}+u_{12}v_{23}\right)
^2-u_{23}v_{12}v_{13}c^{*}+u_{12}v_{13}v_{23}e^{*}\right) , \\ -\frac{
\lambda ^6av_{23}}2\left( cu_{13}v_{12}+\frac{av_{23}}2\right)
,-a^2cu_{13}u_{23}v_{13}v_{23}c^{*}\lambda ^{10}\}\};
\end{array}
\end{equation}
\begin{equation}
\begin{array}{c}
{\tilde N}_{222}^\omega =\{\{0,
\frac{u_{12}^2\lambda ^2}2\left( 1-\lambda ^2\right) , \\ 
\frac{\lambda ^6a}2
\left( a\left( u_{12}u_{23}-v_{12}v_{23}\right)
^2+u_{13}v_{12}v_{23}c-u_{13}v_{12}v_{23}e^{*}\right) \}, \\ 
\{\frac{u_{12}^2\lambda ^2}2\left( 1-\lambda ^2\right) ,0,
\frac{a^2u_{23}^2\lambda ^4}2 \}, \\ 
\{\frac{\lambda ^6a}2\left( a\left( u_{12}u_{23}-v_{12}v_{23}\right)
^2+u_{13}v_{12}v_{23}c-u_{13}v_{12}v_{23}e^{*}\right) , \\ 
\frac{a^2u_{23}^2\lambda ^4}2 ,0\}\} \\ Sp\left( N_{222}^\omega \right) =
\frac 12+\left( v_{12}^2-1\right) \lambda ^2+ \\ \left( \frac 34+\frac{
u_{12}^4}2-v_{12}^2+u_{12}^2v_{12}^2+\frac{v_{12}^4}2+a^2v_{23}^2\right)
\lambda ^4,
\end{array}
\end{equation}
and the flavor antiparticle number fluctuations, $\bar N_{222}(t)$, are 
given by  
\begin{equation}
\label{eq09}
\begin{array}{c}
\bar N_{222}^\Omega =\{\{-eu_{12}u_{13}v_{12}v_{13}e^{*}\lambda ^8,-\frac{
v_{12}^2\lambda ^2}2\left( 1-\lambda ^2\right) ,\frac{\lambda ^6a}2
v_{13}\left( -u_{23}v_{12}c^{*}+u_{12}v_{23}e^{*}\right) \}, \\ \{-
\frac{v_{12}^2\lambda ^2}2\left( 1-\lambda ^2\right) ,0,-\frac{
a^2v_{23}^2\lambda ^4}2\},\{\frac{\lambda ^6a}2v_{13}\left(
-u_{23}v_{12}c^{*}+u_{12}v_{23}e^{*}\right) ,-\frac{a^2v_{23}^2\lambda ^4}2,
\\ -a^2cc^{*}u_{13}u_{23}v_{13}v_{23}\lambda ^{10}\}\};
\end{array}
\end{equation}
\begin{equation}
\label{eq10}
\begin{array}{c}
{\tilde {\bar N}}_{222}^\omega =\{\{0,-\frac{aeu_{12}v_{13}v_{23}}2\lambda 
^6,\frac{ au_{13}v_{12}v_{23}\lambda ^6}2\left( c-e^{*}\right) \}, \\ \{-
\frac{ae^{*}u_{12}v_{13}v_{23}}2\lambda ^6,0,\frac{au_{23}v_{12}v_{13}c^{*}
\lambda ^6}2\}, \\ \{
\frac{au_{13}v_{12}v_{23}\lambda ^6}2\left( c^{*}-e\right) ,\frac{
au_{23}v_{12}v_{13}c\lambda ^6}2,0\}\}; \\ Sp\left( \bar N_{222}^\omega
\right) =v_{12}^2\lambda ^2+\left( a^2v_{23}^2-v_{12}^2\right) \lambda ^4.
\end{array}
\end{equation}

Again in the same initial condition, the mixing quantities for the 
particle of sort \#3 are as follows.
The free-field particle condensate is 
\begin{equation}
Z_1^{\prime }=a^2v_{23}^2\lambda ^4
\end{equation}
and the flavor particle condensate is 
\begin{equation}
\label{eq11}
\begin{array}{c}
Z_{232}^\Omega =\{\{-eu_{12}u_{13}v_{12}v_{13}e^{*}\lambda ^8,
\frac{\lambda ^6av_{12}}2\left( u_{13}v_{23}e+u_{23}v_{13}e^{*}\right) , \\ 
\frac{\lambda ^6v_{13}e^{*}}2\left( au_{12}v_{23}-v_{13}c^{*}\right) \},\{
\frac{\lambda ^6av_{12}}2\left( u_{13}v_{23}e+u_{23}v_{13}e^{*}\right) , \\ 
a^2u_{12}u_{23}v_{12}v_{23}\lambda ^6,-
\frac{\lambda ^4a^2v_{23}^2}2 \} \\ 
\{\frac{\lambda ^6v_{13}e^{*}}2
\left( au_{12}v_{23}-v_{13}c^{*}\right) ,-\frac{\lambda ^4a^2v_{23}^2}2 ,0\}\}
\end{array}
\end{equation}
\begin{equation}
\label{eq12}
\begin{array}{c}
{\tilde Z}_{232}^\omega =\{\{0,a^2eu_{13}u_{23}v_{13}v_{23}e^{*}\lambda 
^{10},- \frac{au_{13}v_{12}v_{23}e^{*}}2\lambda ^6\}, \\ 
\{a^2eu_{13}u_{23}v_{13}v_{23}e^{*}\lambda ^{10},0,-
\frac{au_{23}v_{12}v_{13}c^{*}}2\lambda ^6\}, \\ \{-
\frac{au_{13}v_{12}v_{23}e}2\lambda ^6,-\frac{au_{23}v_{12}v_{13}c}2\lambda
^6,0\}\}; \\ Sp\left( Z_{232}^\omega \right) =a^2v_{23}^2\lambda ^4.
\end{array}
\end{equation}
The flavor particle number fluctuations, $N_{232}(t)=\left|
F_{32}(t)\right| ^2+Z_3(t)$, are given by 
\begin{equation}
\begin{array}{c}
N_{232}^\Omega =\{\{-\lambda ^4u_{12}^2v_{12}^2,-\lambda ^2
\frac{v_{12}^2}2\left( 1-\lambda ^2\right) , \\ 
\frac{\lambda ^6a}2\left(
2av_{12}v_{23}u_{12}u_{23}-u_{23}v_{12}v_{13}c^{*}+v_{13}\left(
u_{23}v_{12}+u_{12}v_{23}\right) e^{*}\right) \}, \\ 
\{-\lambda ^2
\frac{v_{12}^2}2\left( 1-\lambda ^2\right) ,0,-\frac{a^2v_{23}^2}2\lambda ^4\}, \\ 
\{\frac{\lambda ^6a}2\left(
2av_{12}v_{23}u_{12}u_{23}-u_{23}v_{12}v_{13}c^{*}+v_{13}\left(
u_{23}v_{12}+u_{12}v_{23}\right) e^{*}\right) , \\ 
-\frac{a^2v_{23}^2}2\lambda ^4,-a^4u_{23}^2v_{23}^2\lambda ^8\}\};
\end{array}
\end{equation}
\begin{equation}
\begin{array}{c}
{\tilde N}_{232}^\omega =\{\{0,-
\frac{au_{12}\lambda ^6}2\left( v_{13}v_{23}e+u_{13}u_{23}e^{*}\right) , \\ 
\frac{\lambda ^6a}2\left( v_{12}v_{23}\left( u_{13}\left( c-e^{*}\right)
-2au_{12}u_{23}\right) +u_{12}u_{13}u_{23}e^{*}\right) \}, \\ \{-
\frac{au_{12}\lambda ^6}2\left( u_{13}u_{23}e+v_{13}v_{23}e^{*}\right) , \\ 
0,-\frac{a^2u_{23}^2\lambda ^4}2 \}, \\ 
\{\frac{\lambda ^6a}2\left( v_{12}v_{23}\left( u_{13}\left( c^{*}-e\right)
-2au_{12}u_{23}\right) +u_{12}u_{13}u_{23}e\right) , \\ -
\frac{a^2u_{23}^2\lambda ^4}2 ,0\}\}; \\ Sp\left( N_{232}^\omega \right)
=\lambda ^2v_{12}^2+\lambda ^4\left( a^2\left( v_{23}^2+u_{23}^2\right)
-v_{12}^2\left( 1-u_{12}^2\right) \right) .
\end{array}
\end{equation}
Similarly, the flavor antiparticle number fluctuations, $\bar N_{232}(t)$,
are given by  
\begin{equation}
\label{eq13}
\begin{array}{c}
\bar N_{232}^\Omega =\{\{-u_{12}^2v_{12}^2\lambda ^4,-\frac{v_{12}^2\lambda
^2}2\left( 1-\lambda ^2\right) ,\frac{\lambda ^6av_{12}u_{23}}2\left(
2av_{23}u_{12}-v_{13}c^{*}\right) \}, \\ \{-
\frac{v_{12}^2\lambda ^2}2\left( 1-\lambda ^2\right) ,0,-\frac{
acu_{13}v_{12}v_{23}\lambda ^6}2\}, \\ \{\frac{\lambda ^6av_{12}u_{23}}2
\left( 2av_{23}u_{12}-v_{13}c^{*}\right) ,-\frac{acu_{13}v_{12}v_{23}\lambda
^6}2,-a^4u_{23}^2v_{23}^2\lambda ^8\}\};
\end{array}
\end{equation}
\begin{equation}
\label{eq14}
\begin{array}{c}
{\tilde {\bar N}}_{232}^\omega =\{\{0,0,-\frac{av_{12}v_{23}\lambda 
^6}2\left( 2au_{12}u_{23}-u_{13}c\right) \}, \\ \{0,0,
\frac{au_{23}v_{12}v_{13}c^{*}\lambda ^6}2\}, \\ \{-
\frac{av_{12}v_{23}\lambda ^6}2\left( 2au_{12}u_{23}-u_{13}c^{*}\right) ,
\frac{au_{23}v_{12}v_{13}c\lambda ^6}2,0\}\}; \\ Sp\left( \bar N
_{232}^\omega \right) =v_{12}^2\lambda ^2+\lambda ^4v_{12}^2\left(
u_{12}^2-1\right) .
\end{array}
\end{equation}

\subsection{The case 3 boson fields}

We now consider the application to bosons. 
The boson case is not much different from the fermion case.
With the use of
the $\gamma _{ij}^{+},\gamma _{ij}^{-}$ matrices, one can write the ladder
mixing matrices as 
\begin{equation}
\begin{array}{c}
\alpha =\left( 
\begin{array}{ccc}
1-\lambda ^2/2 & \gamma _{12}^{+}\lambda  & \gamma _{13}^{+}A\lambda
^3\left( \rho -i\eta \right)  \\ 
-\gamma _{12}^{+}\lambda  & 1-\lambda ^2/2 & \gamma _{23}^{+}A\lambda ^2 \\ 
\gamma _{13}^{+}A\lambda ^3\left( 1-\rho -i\eta \right)  & -\gamma
_{23}^{+}A\lambda ^2 & 1
\end{array}
\right) ; \\ 
\beta =\left( 
\begin{array}{ccc}
0 & \gamma _{12}^{-}\lambda  & \gamma _{13}^{-}A\lambda ^3\left( \rho -i\eta
\right)  \\ 
\gamma _{12}^{-}\lambda  & 0 & \gamma _{23}^{-}A\lambda ^2 \\ 
-\gamma _{13}^{-}A\lambda ^3\left( 1-\rho -i\eta \right)  & \gamma
_{23}^{-}A\lambda ^2 & 0
\end{array}
\right) .
\end{array}
\end{equation}
We see that $\alpha $ and $\beta $ indeed have the same form as in
the fermion case with the correspondence $\gamma _{ij}^{+}\rightarrow 
u_{ij}$ and $\gamma _{ij}^{-}\rightarrow v_{ij}$: 
\begin{equation}
\begin{array}{c}
\alpha =\left( 
\begin{array}{ccc}
1-\lambda ^2/2 & u_{12}\lambda  & u_{13}c\lambda ^3 \\ 
-u_{12}\lambda  & 1-\lambda ^2/2 & u_{23}a\lambda ^2 \\ 
-u_{13}e\lambda ^3 & -u_{23}a\lambda ^2 & 1
\end{array}
\right) ; \\ 
\beta =\left( 
\begin{array}{ccc}
0 & v_{12}\lambda  & v_{13}c\lambda ^3 \\ 
v_{12}\lambda  & 0 & v_{23}a\lambda ^2 \\ 
v_{13}e\lambda ^3 & v_{23}a\lambda ^2 & 0
\end{array}
\right) .
\end{array}
\end{equation}
This shows that the only difference appears in the quantities that have 
explicit spin dependence, i.e. $F_{\mu\nu}$ and everything involving 
$F_{\mu\nu}$. As a rule, the
quantities for the boson case can be obtained from the fermion 
formulae by simply changing the signs in the terms quadratic in $v_{ij}$. 
We summarize them below.

The oscillation formulas are as follows. For the particle of sort \#1, 
$Q_{212}(t)$ is  
\begin{equation}
\begin{array}{c}
Q_{212}^\Omega =\{\{0,
\frac{\lambda ^2v_{12}^2}2\left( 1-\lambda ^2\right) ,\frac{\lambda
^6av_{13}c^{*}}2\left( u_{23}v_{12}+u_{12}v_{23}\right) \}, \\ \{
\frac{\lambda ^2v_{12}^2}2\left( 1-\lambda ^2\right) ,0,-\frac{\lambda
^6av_{23}}2\left( u_{12}v_{13}c^{*}-u_{13}v_{12}c\right) \}, \\ \{\frac{
\lambda ^6av_{13}c^{*}}2\left( u_{23}v_{12}+u_{12}v_{23}\right) ,-\frac{
\lambda ^6av_{23}}2\left( u_{12}v_{13}c^{*}-u_{13}v_{12}c\right) ,0\}\};
\end{array}
\end{equation}
\begin{equation}
\begin{array}{c}
{\tilde Q}_{212}^\omega =\{\{0,-
\frac{\lambda ^2u_{12}^2}2\left( 1-\lambda ^2\right) ,-\frac{\lambda
^6au_{13}c}2\left( u_{12}u_{23}+v_{12}v_{23}\right) \}, \\ \{-
\frac{\lambda ^2u_{12}^2}2\left( 1-\lambda ^2\right) ,0,\frac{\lambda
^6au_{23}}2\left( u_{12}u_{13}c-v_{12}v_{13}c^{*}\right) \}, \\ \{-
\frac{\lambda ^6au_{13}c^{*}}2\left( u_{12}u_{23}+v_{12}v_{23}\right) ,\frac{
\lambda ^6au_{23}}2\left( u_{12}u_{13}c^{*}-v_{12}v_{13}c\right) ,0\}\}; \\ 
Sp\left( Q_{212}^\omega \right) =\lambda ^2\left( u_{12}^2-v_{12}^2\right)
\left( 1-\lambda ^2\right) .
\end{array}
\end{equation}
For the particle of sort \#2, $Q_{222}(t)$ is 
\begin{equation}
\begin{array}{c}
Q_{222}^\Omega =\{\{0,-
\frac{v_{12}^2\lambda ^2}2\left( 1-\lambda ^2\right) ,-\frac{a^2\left(
u_{23}v_{12}+u_{12}v_{23}\right) ^2}2\lambda ^6\}, \\ \{-
\frac{v_{12}^2\lambda ^2}2\left( 1-\lambda ^2\right) ,0,-\frac{
a^2v_{23}^2\lambda ^4}2\left( 1-\lambda ^2\right) \}, \\ \{-\frac{a^2\left(
u_{23}v_{12}+u_{12}v_{23}\right) ^2}2\lambda ^6,-\frac{a^2v_{23}^2\lambda ^4}
2\left( 1-\lambda ^2\right) ,0\}\};
\end{array}
\end{equation}
\begin{equation}
\begin{array}{c}
{\tilde Q}_{222}^\omega =\{\{0,
\frac{u_{12}^2\lambda ^2}2\left( 1-\lambda ^2\right) ,\frac{a^2\left(
u_{23}u_{12}+v_{12}v_{23}\right) ^2}2\lambda ^6\}, \\ \{
\frac{u_{12}^2\lambda ^2}2\left( 1-\lambda ^2\right) ,0,\frac{
a^2u_{23}^2\lambda ^4}2\left( 1-\lambda ^2\right) \}, \\ \{
\frac{a^2\left( u_{23}u_{12}+v_{12}v_{23}\right) ^2}2\lambda ^6,\frac{
a^2u_{23}^2\lambda ^4}2\left( 1-\lambda ^2\right) ,0\}\}; \\ Sp\left(
Q_{222}^\omega \right) =\frac 12-\lambda ^2+\frac 14\left(
3+2u_{12}^4-4u_{12}^2v_{12}^2+2v_{12}^4\right) \lambda ^4.
\end{array}
\end{equation}
For the particle of sort \#3, $Q_{232}(t)$ is  
\begin{equation}
\begin{array}{c}
Q_{232}^\Omega =\{\{0,-
\frac{av_{12}\lambda ^6}2\left( eu_{13}v_{23}-u_{23}v_{13}e^{*}\right) ,-
\frac{\lambda ^6av_{13}e^{*}}2\left( u_{23}v_{12}+u_{12}v_{23}\right) \}, \\ 
\{-
\frac{av_{12}\lambda ^6}2\left( eu_{13}v_{23}-u_{23}v_{13}e^{*}\right) ,0,
\frac{a^2v_{23}^2\lambda ^4}2-\frac{a^2v_{23}^2\lambda ^6}4\}, \\ \{-\frac{
\lambda ^6av_{13}e^{*}}2\left( u_{23}v_{12}+u_{12}v_{23}\right) ,\frac{
a^2v_{23}^2\lambda ^4}2-\frac{a^2v_{23}^2\lambda ^6}4,0\}\};
\end{array}
\end{equation}
\begin{equation}
\begin{array}{c}
{\tilde Q}_{232}^\omega =\{\{0,
\frac{au_{12}\lambda ^6}2\left( ev_{13}v_{23}-u_{13}u_{23}e^{*}\right) ,
\frac{\lambda ^6au_{13}e^{*}}2\left( u_{12}u_{23}+v_{12}v_{23}\right) \}, \\ 
\{
\frac{au_{12}\lambda ^6}2\left( e^{*}v_{13}v_{23}-u_{13}u_{23}e\right) ,0,-
\frac{a^2u_{23}^2\lambda ^4}2+\frac{a^2u_{23}^2\lambda ^6}4\}, \\ \{
\frac{\lambda ^6au_{13}e}2\left( u_{12}u_{23}+v_{12}v_{23}\right) ,-\frac{
a^2u_{23}^2\lambda ^4}2+\frac{a^2u_{23}^2\lambda ^6}4,0\}\}; \\ Sp\left(
Q_{232}^\omega \right) =a^2\left( u_{23}^2-v_{23}^2\right) \lambda ^4\left(
1-\lambda ^2/2\right) .
\end{array}
\end{equation}

The non-equal time commutators are given by: 
\begin{equation}
\begin{array}{c}
F_{11}(t)=e^{-i\epsilon _1t}+\lambda ^2\left( -e^{-i\epsilon
_1t}+u_{12}^2e^{-i\epsilon _2t}-v_{12}^2e^{i\epsilon _2t}\right) , \\ 
F_{12}(t)=F_{21}(t)=\lambda u_{12}\left( e^{-i\epsilon
_2t}-e^{-i\epsilon _1t}\right) +\lambda ^3
\frac{u_{12}}2\left( e^{-i\epsilon _1t}-e^{-i\epsilon _2t}\right)  \\ 
F_{13}(t)=(F_{31}(-t))^{*}=\lambda ^3\left( cu_{13}e^{-i\epsilon
_3t}-au_{12}u_{23}e^{-i\epsilon _2t}-av_{12}v_{23}e^{i\epsilon
_2t}-e^{*}u_{13}e^{-i\epsilon _1t}\right) , \\ 
F_{22}(t)=e^{-i\epsilon _2t}+\lambda ^2\left( -e^{-i\epsilon
_2t}+u_{12}^2e^{-i\epsilon _1t}-v_{12}^2e^{i\epsilon _1t}\right) , \\ 
F_{23}(t)=F_{32}(t)=\lambda ^2au_{23}\left( e^{-i\epsilon
_3t}-e^{-i\epsilon _2t}\right) , \\ 
F_{33}(t)=e^{-i\epsilon _3t};
\end{array}
\end{equation}
\begin{equation}
\begin{array}{c}
G_{11}(t)=\lambda ^2u_{12}v_{12}\left( e^{-i\epsilon _2t}-e^{i\epsilon
_2t}\right) , \\ 
G_{12}(t)=-(G_{21}(t))^{*}=\lambda v_{12}\left( e^{-i\epsilon
_1t}-e^{i\epsilon _2t}\right) +\lambda ^3
\frac{v_{12}}2\left( -e^{-i\epsilon _1t}+e^{i\epsilon _2t}\right) , \\ 
G_{13}(t)=-(G_{31}(t))^{*}=\lambda ^3\left( au_{23}v_{12}e^{i\epsilon
_2t}+ev_{13}\cdot e^{-i\epsilon _1t}+au_{12}v_{23}e^{-i\epsilon
_2t}-v_{13}c^{*}e^{i\epsilon _3t}\right) , \\ 
G_{22}(t)=\lambda ^2u_{12}v_{12}\left( -e^{-i\epsilon _1t}+e^{i\epsilon
_1t}\right) , \\ 
G_{23}(t)=-(G_{32}(t))^{*}=\lambda ^2av_{23}\left( e^{-i\epsilon
_2t}-e^{i\epsilon _3t}\right) , \\ 
G_{33}(t)=\lambda ^4a^2u_{23}v_{23}\left( e^{i\epsilon _2t}-e^{-i\epsilon
_2t}\right) .
\end{array}
\end{equation}
The vacuum structure is given by the fermion $\hat Z$ (Eq.(\ref{eq02})) 
with the normalization constant 
$$
{\cal Z}\approx 1+v_{12}^2\lambda
^2+(v_{12}^2+a^2v_{23}^2+v_{12}^4-v_{12}^2u_{12}^2)\lambda ^4+\ldots 
$$
If the particle of sort \#2 was emitted initially, then the particle of 
sort \#1 has the following free-field particle condensate: 
\begin{equation}
Z_1^{\prime }=v_{12}^2\lambda ^2
\end{equation}
and the flavor particle condensate identical to the fermion
case,{\it i.e.} Eqs.(\ref{eq03}) and (\ref{eq04}).

The flavor particle number fluctuation $N_{212}(t)$ is given by  
\begin{equation}
\begin{array}{c}
N_{212}^\Omega =\{\{-\lambda ^4u_{12}^2v_{12}^2,-\lambda ^2
\frac{v_{12}^2}2\left( 1-\lambda ^2\right) , \\ 
\frac{\lambda^6a}{2}\left( 2au_{12}u_{23}v_{12}v_{23}+u_{12}v_{13}v_{23}(e^*+c^*)-
u_{23}v_{12}v_{13}c^* \right) \}, \\ 
\{-\lambda ^2
\frac{v_{12}^2}2\left( 1-\lambda ^2\right) ,0,-\frac{a^2v_{23}^2\lambda ^4}2 \}, \\ 
\{\frac{\lambda^6a}{2}\left( 2au_{12}u_{23}v_{12}v_{23}+u_{12}v_{13}v_{23}(e^*+c^*)-
u_{23}v_{12}v_{13}c^* \right),
\\ -\frac{a^2v_{23}^2\lambda ^4}2 ,-a^4u_{23}^2v_{23}^2\lambda ^8\}\};
\end{array}
\end{equation}
\begin{equation}
\begin{array}{c}
{\tilde N}_{212}^\omega =\{\{0,-\lambda ^2u_{12}^2\left( 1-\lambda ^2\right) 
, \\ -\frac{\lambda ^6a}2\left( v_{12}v_{23}\left(
u_{13}e^{*}-u_{13}c+2au_{12}u_{23}\right) +cu_{12}u_{13}u_{23}\right)
\},\{-\lambda ^2u_{12}^2\left( 1-\lambda ^2\right) , \\ 
0, \frac{\lambda ^6au_{23}}2\left( cu_{12}u_{13}+c^{*}v_{12}v_{13}\right) \},
\\ \{-
\frac{\lambda ^6a}2\left( v_{12}v_{23}\left(
u_{13}e-u_{13}c^{*}+2au_{12}u_{23}\right) +c^{*}u_{12}u_{13}u_{23}\right) ,
\\ \frac{\lambda ^6au_{23}}2\left( c^{*}u_{12}u_{13}+cv_{12}v_{13}\right)
,0\}\}; \\ Sp\left( N_{212}^\omega \right) =\lambda ^2\left(
u_{12}^2+v_{12}^2\right) +\lambda ^4\left(
a^2v_{23}^2-v_{12}^2-u_{12}^2\left( 1-v_{12}^2\right) \right) .
\end{array}
\end{equation}

The flavor antiparticle number fluctuation $\bar N_{212}(t)$ is given by 
\begin{equation}
\begin{array}{c}
\bar N_{212}^\Omega =\{\{-u_{12}^2v_{12}^2\lambda ^4,-v_{12}^2\lambda
^2\left( 1-\lambda ^2\right) , \\ 
\frac{\lambda ^6a}2\left(
-2u_{23}v_{12}v_{13}c^{*}+u_{12}v_{23}\left(
2au_{23}v_{12}+v_{13}e^{*}\right) \right) \}, \\ 
\{-v_{12}^2\lambda ^2\left(1-\lambda ^2\right) ,0,
-\frac{a^2v_{23}^2\lambda ^4}2 \}, \\ \{
\frac{\lambda ^6a}2\left( -2u_{23}v_{12}v_{13}c^{*}+u_{12}v_{23}\left(
2au_{23}v_{12}+v_{13}e^{*}\right) \right) , \\ 
-\frac{a^2v_{23}^2\lambda ^4}2,-a^4u_{23}^2v_{23}^2\lambda ^8\}\};
\end{array}
\end{equation}
\begin{equation}
\begin{array}{c}
{\tilde {\bar N}}_{212}^\omega =\{\{0,-\frac{aeu_{12}v_{13}v_{23}\lambda 
^6}2,-\frac{
av_{12}v_{23}\lambda ^6}2\left( -2cu_{13+}2au_{12}u_{23}+u_{13}e^{*}\right)
\}, \\ 
\{-\frac{ae^{*}u_{12}v_{13}v_{23}\lambda ^6}2,0,au_{23}v_{12}v_{13}c^{*}\lambda
^6\}, \\ 
\{-\frac{av_{12}v_{23}\lambda ^6}2\left(
-2c^{*}u_{13+}2au_{12}u_{23}+u_{13}e\right) ,au_{23}v_{12}v_{13}c\lambda
^6,0\}\}; \\ Sp\left( \bar N_{212}^\omega \right) =2v_{12}^2\lambda
^2+\lambda ^4\left( u_{12}^2v_{12}^2+a^2v_{23}^2-2v_{12}^2\right) .
\end{array}
\end{equation}

For the same initial condition, the particle of sort \#2 has the 
free-field condensate given by: 
\begin{equation}
Z_2^{\prime }=v_{12}^2\lambda ^2+a^2v_{23}^2\lambda ^4
\end{equation}
and the flavor particle condensate identical to Eqs.(\ref
{eq07}) and (\ref{eq08}).

The flavor particle number fluctuation $N_{222}(t)$ is  
\begin{equation}
\begin{array}{c}
N_{222}^\Omega =\{\{-2u_{12}^2v_{12}^2\lambda
^4,-v_{12}^2\lambda ^2\left(
1-\lambda ^2\right) , \\ 
-\frac{\lambda ^6a}2\left( a\left( u_{23}v_{12}-u_{12}v_{23}\right)
^2+u_{23}v_{12}v_{13}c^{*}-u_{12}v_{13}v_{23}e^{*}\right) \}, \\ 
\{-v_{12}^2\lambda ^2\left( 1-\lambda ^2\right) ,0,-a^2v_{23}^2\lambda ^4 \}, \\ 
\{-\frac{\lambda ^6a}2\left( a\left( u_{23}v_{12}-u_{12}v_{23}\right)
^2+u_{23}v_{12}v_{13}c^{*}-u_{12}v_{13}v_{23}e^{*}\right) , \\ 
-a^2v_{23}^2\lambda ^4 ,-2a^4u_{23}^2v_{23}^2\lambda ^8\}\};
\end{array}
\end{equation}
\begin{equation}
\begin{array}{c}
{\tilde N}_{222}^\omega =\{\{0,
\frac{u_{12}^2\lambda ^2}2\left( 1-\lambda ^2\right) , \\ 
\frac{\lambda ^6a}2
\left( cu_{13}v_{12}v_{23}+a\left( u_{12}u_{23}-v_{12}v_{23}\right)
^2-u_{13}v_{12}v_{23}e^{*}\right) \}, \\ 
\{\frac{u_{12}^2\lambda ^2}2\left( 1-\lambda ^2\right) ,0,\frac{
a^2u_{23}^2\lambda ^4}2 \}, \\ 
\{\frac{\lambda ^6a}2\left( c^{*}u_{13}v_{12}v_{23}+a\left(
u_{12}u_{23}-v_{12}v_{23}\right) ^2-u_{13}v_{12}v_{23}e\right) , \\ 
\frac{a^2u_{23}^2\lambda ^4}2 ,0\}\}; \\ 
Sp\left( N_{222}^\omega \right) =
\frac 12+\left( v_{12}^2-1\right) \lambda ^2+ \\ \left( \frac 34+\frac{
u_{12}^4}2-v_{12}^2+u_{12}^2v_{12}^2+\frac{v_{12}^4}2+a^2v_{23}^2\right)
\lambda ^4.
\end{array}
\end{equation}

The flavor antiparticle number fluctuation $\bar N_{222}(t)$ is 
\begin{equation}
\begin{array}{c}
\bar N_{222}^\Omega =\{\{-2u_{12}^2v_{12}^2\lambda ^4,-\frac{v_{12}^2\lambda
^2}2\left( 1-\lambda ^2\right) , \\ \frac{\lambda ^6a}2\left(
-u_{23}v_{12}v_{13}c^{*}+u_{12}v_{23}\left(
4au_{23}v_{12}+v_{13}e^{*}\right) \right) \}, \\ \{-
\frac{v_{12}^2\lambda ^2}2\left( 1-\lambda ^2\right) ,0,-\frac{
a^2v_{23}^2\lambda ^4}2\}, \\ \{
\frac{\lambda ^6a}2\left( -u_{23}v_{12}v_{13}c^{*}+u_{12}v_{23}\left(
4au_{23}v_{12}+v_{13}e^{*}\right) \right) ,-\frac{a^2v_{23}^2\lambda ^4}2,
\\ -2a^4u_{23}^2v_{23}^2\lambda ^8\}\};
\end{array}
\end{equation}
\begin{equation}
\begin{array}{c}
{\tilde {\bar N}}_{222}^\omega =\{\{0,-\frac{aeu_{12}v_{13}v_{23}}2\lambda 
^6,\frac{ av_{12}v_{23}\lambda ^6}2\left( \left( c-e^{*}\right)
u_{13}-4au_{12}u_{23}\right) \}, \\ \{-
\frac{ae^{*}u_{12}v_{13}v_{23}}2\lambda ^6,0,\frac{au_{23}v_{12}v_{13}c^{*}
\lambda ^6}2\}, \\ \{
\frac{av_{12}v_{23}\lambda ^6}2\left( \left( c^{*}-e\right)
u_{13}-4au_{12}u_{23}\right) ,\frac{au_{23}v_{12}v_{13}c\lambda ^6}2,0\}\};
\\ Sp\left( \bar N_{222}^\omega \right) =v_{12}^2\lambda ^2+\left(
a^2v_{23}^2-v_{12}^2+2u_{12}^2v_{12}^2\right) \lambda ^4.
\end{array}
\end{equation}

Finally, for the same initial condtion, the particle of sort \#3  
has the free-field particles condensate given by
\begin{equation}
Z_1^{\prime }=a^2v_{23}^2\lambda ^4
\end{equation}
and the flavor particle condensate identical to Eqs.(\ref
{eq11}) and (\ref{eq12}).

The flavor particle number fluctuation $N_{232}(t)$ is given by 
\begin{equation}
\begin{array}{c}
N_{232}^\Omega =\{\{-\lambda ^4u_{12}^2v_{12}^2,-\lambda ^2
\frac{v_{12}^2}2\left( 1-\lambda ^2\right) , \\ 
\frac{\lambda ^6a}2\left(
-u_{23}v_{12}v_{13}\left( c^{*}+e^{*}\right)
+2au_{12}u_{23}v_{12}v_{23}+e^{*}u_{12}v_{13}v_{23}\right) \}, \\ 
\{-\lambda^2
\frac{v_{12}^2}2\left( 1-\lambda ^2\right) ,0, \\ 
-\frac{a^2v_{23}^2}2\lambda ^4 \}, \\ 
\{\frac{\lambda ^6a}2\left( -u_{23}v_{12}v_{13}\left( c^{*}+e^{*}\right)
+2au_{12}u_{23}v_{12}v_{23}+e^{*}u_{12}v_{13}v_{23}\right) , \\ 
-\frac{a^2v_{23}^2}2\lambda ^4 ,-a^4u_{23}^2v_{23}^2\lambda ^8\}\};
\end{array}
\end{equation}
\begin{equation}
\begin{array}{c}
{\tilde N}_{232}^\omega =\{\{0,-
\frac{au_{12}\lambda ^6}2\left( v_{13}v_{23}e+u_{13}u_{23}e^{*}\right) , \\ 
\frac{\lambda ^6a}2\left( v_{12}v_{23}\left( u_{13}\left( c-e^{*}\right)
-2au_{12}u_{23}\right) +u_{13}u_{12}u_{23}e^{*}\right) \}, \\ 
\{-\frac{au_{12}\lambda ^6}2\left( v_{13}v_{23}e^{*}+u_{13}u_{23}e\right) ,0,
-\frac{a^2u_{23}^2\lambda ^4}2+, \\ 
\{\frac{\lambda ^6a}2\left( v_{12}v_{23}\left( u_{13}\left( c^{*}-e\right)
-2au_{12}u_{23}\right) +u_{13}u_{12}u_{23}e\right) , \\ 
-\frac{a^2u_{23}^2\lambda ^4}2 ,0\}\}; \\ Sp\left( N_{232}^\omega \right)
=\lambda ^2v_{12}^2+\lambda ^4\left( a^2\left( u_{23}^2+v_{23}^2\right)
-v_{12}^2\left( 1-u_{12}^2\right) \right) .
\end{array}
\end{equation}

The flavor antiparticle number fluctuation $\bar N_{232}(t)$ is given by 
\begin{equation}
\begin{array}{c}
\bar N_{232}^\Omega =\{\{-\lambda ^4u_{12}^2v_{12}^2,-\lambda ^2\frac{
v_{12}^2}2\left( 1-\lambda ^2\right) , \\ 
\frac{\lambda ^6a}2\left(
-u_{23}v_{12}v_{13}c^{*}+2au_{12}u_{23}v_{12}v_{23}+2e^{*}u_{12}v_{13}v_{23}
\right) \}, \\ 
\{-\lambda ^2\frac{v_{12}^2}2\left( 1-\lambda ^2\right) ,0,
-a^2v_{23}^2\lambda ^4 \}, \\ \{
\frac{\lambda ^6a}2\left(
-u_{23}v_{12}v_{13}c^{*}+2au_{12}u_{23}v_{12}v_{23}+2e^{*}u_{12}v_{13}v_{23}
\right) , \\ 
-a^2v_{23}^2\lambda ^4 ,-a^4u_{23}^2v_{23}^2\lambda ^8\}\};
\end{array}
\end{equation}
\begin{equation}
\begin{array}{c}
{\tilde {\bar N}}_{232}^\omega =\{\{0,-\lambda ^6au_{12}v_{13}v_{23}e, \\ 
\frac{\lambda
^6a}2v_{12}v_{23}\left( u_{13}\left( c-2e^{*}\right) -2au_{12}u_{23}\right)
\}, \\ 
\{-\lambda ^6au_{12}v_{13}v_{23}e^{*},0,
\frac{\lambda ^6au_{23}}2c^{*}v_{12}v_{13}\}, \\ \{
\frac{\lambda ^6a}2v_{12}v_{23}\left( u_{13}\left( c^{*}-2e\right)
-2au_{12}u_{23}\right) ,\frac{\lambda ^6au_{23}}2cv_{12}v_{13},0\}\}; \\ 
Sp\left( N_{232}^\omega \right) =\lambda ^2v_{12}^2+\lambda ^4\left(
2a^2v_{23}^2-v_{12}^2\left( 1-u_{12}^2\right) \right) .
\end{array}
\end{equation}
\end{appendix}

\end{document}